\documentclass[12pt]{article}
\usepackage{epsfig,amsfonts,amssymb}
\usepackage{hyperref}
\usepackage{cite}
\input epsf.sty
\usepackage{cite}

\def\ZZZ{{\hbox{ Z\kern-1.6mm Z}}}
\def\RRR{{\hbox{ R\kern-2.4mm R}}}
\def\CCC{{\hbox{ C\kern-2.0mm C}}}
\def\zzz{{\hbox{z\kern-1mm z}}}

\newcommand{\qeq}{{\hbox{=\kern-2.3mm ? \kern.5mm }}}
\renewcommand{\qeq}{=}

\newcommand{\eps}{\epsilon}

\newcommand{\vp}{\varphi}

\newcommand{\VV}{{\cal V}}
\newcommand{\BB}{{\cal B}}

\newcommand{\AAA}{{\cal A}}
\newcommand{\GG}{{\cal G}}
\newcommand{\KK}{{\cal K}}

\newcommand{\HH}{{\cal H}}
\newcommand{\MM}{{\cal M}}

\newcommand{\OO}{{\cal O}}
\newcommand{\QQ}{{\cal Q}}
\newcommand{\PP}{{\cal P}}

\newcommand{\LL}{{\cal L}}

\newcommand{\wt}{\widetilde}
\newcommand{\wh}{\widehat}

\newcommand{\RR}{{\cal R}}

\newcommand{\be}{\begin{equation}}
\newcommand{\ee}{\end{equation}}
\newcommand{\ben}{\begin{eqnarray}\displaystyle}
\newcommand{\een}{\end{eqnarray}}

\newcommand{\refb}[1]{(\ref{#1})}
\newcommand{\p}{\partial}
\newcommand{\sectiono}[1]{\section{#1}\setcounter{equation}{0}}

\def\one{{\hbox{ 1\kern-.8mm l}}}
\def\zero{{\hbox{ 0\kern-1.5mm 0}}}

\newcommand{\bea}[1]{\begin{eqnarray}\label{#1} }
\newcommand{\eea}{\end{eqnarray}}

\newcommand{\eqref}{\refb}

\newcommand{\XX}{{\cal X}}

\usepackage{bm}
\usepackage[table]{xcolor}

\def\PSZ{\Psi_0}

\begin{document}

\baselineskip 24pt

\begin{center}
{\Large \bf  Background Independence of  Closed Superstring Field Theory}

\end{center}

\vskip .6cm
\medskip

\vspace*{4.0ex}

\baselineskip=18pt

\centerline{\large \rm Ashoke Sen}

\vspace*{4.0ex}

\centerline{\large \it Harish-Chandra Research Institute, HBNI}
\centerline{\large \it  Chhatnag Road, Jhusi,
Allahabad 211019, India}

\vspace*{1.0ex}
\centerline{\small E-mail:  sen@mri.ernet.in}

\vspace*{5.0ex}

\centerline{\bf Abstract} \bigskip

Given a family of world-sheet superconformal field theories related by marginal
deformation, we can formulate superstring field theory based on any of these world-sheet 
theories. Background independence is the statement that these different superstring field
theories are related to each other  by field redefinition. We prove background independence of
closed superstring field theory.

\vfill \eject

\baselineskip=18pt

\tableofcontents

\sectiono{Introduction} \label{sintro}

Dealing with the problem of mass renormalization and vacuum shift in superstring perturbation
theory requires formulation of quantum superstring field theory. We now have such a field theory for
heterotic and type II string theories (for a review see \cite{1703.06410}) arising out of a generalization of the
bosonic closed string field theory constructed in \cite{saadi, kugo,9206084} and tree level Neveu-Schwarz (NS) sector
superstring
field theory described in \cite{9202087}.
However this field theory is apparently background dependent.
One starts with a specific world-sheet superconformal field theory (SCFT)
describing a specific string
compactification, and uses the correlation functions of vertex operators in this world-sheet theory
to construct the interaction vertices of the superstring field theory. Therefore two different world-sheet
theories describing different string compactifications will lead to apparently different superstring field 
theories. Background independence is the statement that these apparently different superstring field
theories are related to each other by field redefinition. Our goal in this paper will be to prove background
independence for the case where the different world-sheet theories are related by marginal 
deformation. Besides being of theoretical interest this also has `practical' consequences -- for example
background independence of superstring field theory is used implicitly in the 
proof of soft theorem (see {\it e.g.} \cite{1703.00024})
where it is assumed that the response of string field theory to an infinitesimal on-shell graviton field
is identical to that under a change in the target space metric in the underlying world-sheet theory.

Background independence of bosonic string field theory was proved in \cite{back1,back2}. As in those
papers, we shall restrict our analysis to the case where the pair of world-sheet theories, for which we want
to establish the equivalence of the corresponding string field theories, are related to each other by
infinitesimal marginal deformation. Once the result is proved for infinitesimal deformations, it also
establishes the result for finite marginal deformations since a finite marginal deformation can be built
from successive application of infinitesimal deformations. Of course, beyond a critical distance between 
the world-sheet theories
the required field redefinition  may diverge,  
but this is simply a reflection of the fact that the coordinate
system in the space of string fields formulated around one world-sheet theory may break down beyond
a certain distance.

Superstring field theory enjoys infinite parameter gauge invariance. Due to this gauge invariance 
the field redefinition that relates a pair of superstring field theories is not unique -- given one such field
redefinition we can find infinite number of other field redefinitions that differ from the first one by gauge
transformation with possibly field dependent gauge transformation parameters. This will be seen
explicitly in our analysis.

Our analysis will differ from that of \cite{back2} for bosonic string field theory in one important way. The
analysis of \cite{back2} focussed on the Batalin-Vilkovisky (BV) master action. For this reason establishing
equivalence of two superstring field theories required taking into account possible change in the
integration measure under the field redefinition. Here we shall work with the one particle irreducible (1PI)
effective action. Since the tree amplitudes computed from the 1PI effective action give the full amplitudes
of string theory, we do not need to worry about the change in the integration measure. In fact it is also
not necessary to prove the equivalence of the actions -- it is sufficient to show that the equations of motion
of the two theories get related to each other by field redefinition. This is useful due to the fact that 
superstring field theory, as formulated in \cite{1703.06410}, contains a set of additional free 
fields besides the 
interacting string field. While these 
additional fields are needed for writing down the action, 
the interacting field equations can be written in terms of the physical string fields, and contains
full information about the S-matrix. Therefore it will be enough to show that the interacting field
equations in the two theories are related to each other by field redefinition. This is the strategy
we shall follow.

Rest of the paper is organized as follows. In section \ref{srev} we review some of the details of the
superstring field theory and also introduce some new notations that will simplify our analysis.
In section \ref{s3} we discuss two ways of describing string field theory around a new background. 
The first corresponds to deforming the world-sheet SCFT by an infinitesimal
marginal deformation and formulating string field theory around the new background. The second approach
is to take the original string field theory action and expand it around an infinitesimal 
classical solution to the linearized equations of 
motion corresponding to the same marginal deformation. 
Background independence is the statement that these two string field theories are related
by a field redefinition. 
In section \ref{sgeo} we give geometric interpretations of the kinetic and interaction
terms of the superstring field theory around marginally deformed background that 
is needed for our proof of background independence. Using these results, 
we give an explicit proof of the background independence in 
section \ref{s4} 
by
describing a systematic algorithm for constructing the field redefinition that relates
the two versions of superstring field theory. In section 
\ref{sgen} we describe extension of our analysis to type II string theories and
also possible applications of our method to other versions of open and
closed superstring field theories.

\sectiono{Review of superstring field theory and some notations} \label{srev}

In this section we shall first review the construction of superstring field theory described in
\cite{1703.06410} and then introduce some new notations that will simplify our analysis.

\subsection{Superstring field theory} \label{srevssft}

We begin with 
a very brief review of superstring field theory -- more details can be found in \cite{1703.06410}.
For simplicity we shall consider heterotic string theory, but the
analysis can be easily generalized to type II string theory. 
Our starting point is the 
matter-ghost world-sheet superconformal
field theory, describing string theory in a specific background.
We denote by $b_n,c_n, L_n$ and $\bar b_n, \bar c_n, \bar L_n$ the usual modes
of the holomorphic and anti-holomorphic parts of the anti-commuting ghost fields and
the total stress tensor. We also denote by $\beta,\gamma$ the superconformal ghosts
and by $\xi,\eta,\phi$ the fields obtained via bosonization of these fields:
\be
\beta = \p\xi\, e^{-\phi}, \quad \gamma=\eta\, e^\phi
\ee
satisfying the operator product expansion 
\ben
&& \xi(z) \, \eta(w) =(z-w)^{-1} + \hbox{non-singular terms}, \nonumber \\
&& 
e^{q\phi}(z) \, e^{q'\phi}(w) = (z-w)^{-q\, q'} e^{(q+q')\phi}(w)+ \hbox{less singular terms}\, .
\een
$\xi_n,\eta_n$ will denote the modes of the $\xi,\eta$ fields.
We define the
picture number of a state such that $\xi$ carries picture number 
$1$, $\eta$ carries picture number $-1$, $e^{q\phi}$ carries picture number
$q$ and other fields carry picture number 0.
We shall denote by
$\HH_p$ the Hilbert space of the world-sheet 
theory carrying picture number $p$ and
subject to the following 
constraints on the states
\ben
&& b_0^- |\Psi\rangle = 0, \quad L_0^- |\Psi\rangle = 0, \quad \eta_0|\Psi\rangle=0, 
\nonumber \\
&& b_0^\pm 
\equiv b_0\pm \bar b_0, \quad L_0^\pm \equiv L_0\pm \bar L_0, \quad 
c_0^\pm \equiv {1\over 2} (c_0\pm \bar c_0)\, .
\een
States in $\HH_p$ for $p\in\ZZZ$ are NS sector states and for $p\in\ZZZ+{1\over 2}$ are
Ramond (R) sector states.

$Q_B$ will denote the nilpotent BRST operator
\be \label{exx1}
Q_B = \ointop dz \jmath_B(z) + \ointop d\bar z \bar \jmath_B(\bar z)\, ,
\ee
where
\be\label{exx2}
\bar \jmath_B(\bar z) = \bar c(\bar z) \bar T_m(\bar z)
+\bar b(\bar z) \bar c(\bar z) \bar\p \bar c(\bar z)\, ,
\ee
\be \label{exx3}
\jmath_B(z) =c(z) (T_{m}(z) + T_{\beta,\gamma}(z) )+ \gamma (z) T_F(z) 
+ b(z) c(z) \p c(z) 
-{1\over 4} \gamma(z)^2 b(z)\, ,
\ee
and $\ointop$ is normalized so that $\ointop dz/z=1$, $\ointop d\bar z/\bar z=1$.
$T_m$ denotes the stress tensor of the matter SCFT, $T_{\beta,\gamma}$ denotes the
stress tensor of the $\beta,\gamma$ system or equivalently the $\xi,\eta,\phi$ system
and $T_F$ denotes the supercurrent of the matter SCFT.

The 
picture changing
operator (PCO) $\XX$
is defined as\cite{FMS,Verlinde:1987sd}
\be \label{epicture}
\XX(z) = \{Q_B, \xi(z)\} = c \, \partial \xi + 
e^\phi T_F - {1\over 4} \p \eta \, e^{2\phi} \, b
- {1\over 4} \p\left(\eta \, e^{2\phi} \, b\right)\, .
\ee
This is a BRST invariant primary operator of dimension zero
carrying picture number $1$.
$\XX_0$ will denote the zero mode of $\XX(z)$:
\be
\XX_0=\int {dz\over z} \, \XX(z)\,  .
\ee
Also we define the operator $\GG$ such that, acting on a state in $\HH_p$,
\be\label{edefgg}
\GG=
\cases{
1 \quad \hbox{if} \quad p\in \ZZZ, \cr
\XX_0 \quad \hbox{if} \quad p\in \ZZZ+{1\over 2}\, . 
}
\ee

Heterotic string field theory contains a pair of fields $|\Psi\rangle$ and $|\wt\Psi\rangle$
which are taken to be arbitrary states in 
\be \label{edefhatwt}
\wh\HH\equiv 
\HH_{-1}+\HH_{-1/2}, \quad \hbox{and} \quad
\wt\HH\equiv \HH_{-1}+\HH_{-3/2}\, ,
\ee 
respectively. 
If $\{|\hat\vp_r\rangle\}$ 
denote the basis states in $\wh\HH$ 
then we expand the string field $|\Psi\rangle$ as
a linear combination $\sum_r \psi_r |\hat\vp_r\rangle$ and identify the coefficients of expansion 
$\psi_r$ as
the dynamical variables of the theory. 
One of the key features of closed string field is that the coefficient $\psi_r$ has the same grassmann
parity as the vertex operator $\hat\vp_r$ of the SCFT so that $\Psi$ is always grassmann even.
Similar remarks hold for $|\wt\Psi\rangle$.

The string field action $S$ is a function of these variables $\psi_r$ and $\wt\psi_r$. 
For our purpose it will be most convenient to work with the
one particle irreducible (1PI) effective action which takes the form:\footnote{The
genus zero contribution to $\{\Psi^N\}$ vanishes for $N=1,2$.}
\be \label{eaction}
S = g_s^{-2} \left[ -{1\over 2} \langle \wt\Psi| c_0^- \GG \, Q_B |\wt\Psi\rangle
+ \langle \wt\Psi| c_0^-  Q_B |\Psi\rangle + \sum_{N\ge 1} {1\over N!} \{\Psi^N\}\right]\, .
\ee
Here $g_s$ is the string coupling.
The last term involving $\{\Psi^N\}$ describes interaction term and is defined as
follows. 
\begin{enumerate}
\item We denote by $\MM_{g,m,n}$ the moduli space of genus $g$ Riemann surfaces
carrying $m$  NS punctures and $n$ R 
punctures. We include in the definition of $\MM_{g,m,n}$ the choice of
spin structure.
\item
We define $\wt\PP_{g,m,n}$ to be 
a fiber bundle with base $\MM_{g,m,n}$ and fiber containing
information on the choice of local coordinates at each puncture and the locations
of $2g-2+m + n/2$ PCO's on the Riemann surface. Therefore specifying a section
of $\wt\PP_{g,m,n}$ corresponds to making a specific choice of local coordinates at
the punctures and choice of PCO locations for every Riemann surface of genus $g$ and
$m+n$ punctures. 
\item We shall use the name generalized subspace to denote a formal
weighed sum of subspaces of
$\wt\PP_{g,m,n}$ with the understanding that integration over such weighted subspaces
mean weighted sum of integrals over each of the subspaces that are being summed.
The generalized subspaces 
can include vertical segments along
which the PCO locations change from an initial configuration to a final configuration 
one at a time, keeping the moduli and local coordinates at the punctures fixed.
These vertical segments can pass through spurious 
poles\cite{Verlinde:1987sd,lechtenfeld,morozov} where the conformal field theory
correlation functions on the corresponding Riemann surfaces have singularities.
However all the relevant integrals along
these vertical segments will be well defined\cite{1408.0571,1504.00609,1710.07232} , 
and as a result we can continue to treat the
vertical segments as regular subspaces of $\wt\PP_{g,m,n}$ for various manipulations.
\item We shall use the name `section segment' 
of $\wt\PP_{g,m,n}$ to denote a $6g-6+2m+2n$ dimensional generalized subspace of 
$\wt\PP_{g,m,n}$, such that the projection of its boundary on the base of $\wt\PP_{g,m,n}$
encloses a proper subspace of the moduli space $\MM_{g,m,n}$. \item For given $g,m,n$
and a set of $m$ different NS 
states $K_1,\cdots K_m$ and $n$ different R states $L_1,\cdots L_n$, one can define a natural
$p$-form $\Omega^{(g,m,n)}_p(K_1,\cdots K_m,L_1,\cdots, L_n)$ on $\wt\PP_{g,,m,n}$
for any integer $p$.
At a given point $\sigma$ in $\wt\PP_{g,,m,n}$, the expression for 
$\Omega^{(g,m,n)}_p(K_1,\cdots K_m,L_1,\cdots, L_n)$
involves correlation function of the vertex operators $\{K_i\}, \{L_i\}$ inserted using the 
appropriate local coordinate system at the punctures corresponding to the point $\sigma$, 
PCO's inserted at appropriate locations corresponding to the point $\sigma$,
and other insertions of ghost fields. 
\item 
We now define
\ben \label{edefcurly}
&&
\{K_1\cdots K_m\, L_1 \cdots L_n\}_g=
\int_{\RR_{g,m,n}} \, \Omega^{(g,m,n)}_{6g-6+2m+2n}(K_1,\cdots K_m,L_1,\cdots, L_n), \nonumber \\
&& \{K_1\cdots K_m\, L_1 \cdots L_n\} = \sum_{g=0}^\infty (g_s)^{2g} \,
\{K_1\cdots K_m\, L_1 \cdots L_n\}_g \, ,
\een
where $\RR_{g,m,n}$ is a section segment of $\wt\PP_{g,m,n}$  that
never includes singular Riemann surfaces corresponding to separating type degeneration but includes
singular Riemann surfaces corresponding to non-separating type 
degenerations.
$\RR_{g,m,n}$ is taken to be symmetric under the exchange of any pair of NS punctures or any
pair of R punctures, and satisfies some other constraints 
that we shall describe shortly. 
\item Given the definition of $\{\cdots\}$ in \refb{edefcurly},
the interaction  term $\{\Psi^N\}$ in \refb{eaction} 
just means that the state $\Psi$ is inserted $N$ times inside the curly bracket.
\end{enumerate}

$\Omega^{(g,m,n)}_p(K_1,\cdots K_m,L_1,\cdots, L_n)$ satisfies the useful identity
\ben \label{euse}
&& \Omega^{(g,m,n)}_p(Q_B \, K_1,\cdots K_m,L_1,\cdots, L_n)
+\cdots + \Omega^{(g,m,n)}_p(K_1,\cdots K_m,L_1,\cdots, Q_B L_n) \nonumber \\
&=& (-1)^p d\Omega^{(g,m,n)}_{p-1}(K_1,\cdots K_m,L_1,\cdots, L_n)\, .
\een
In writing this identity we have assumed that if any of the vertex operators $K_i$ or $L_j$
are grassmann odd, they have been multiplied by a grassmann odd c-number to make
them grassmann even. We can recover the correct sign of each term by pulling these
grassmann odd c-numbers in each term to the extreme right\footnote{The reason that it is
more convenient to move the grassmann odd parameters to the extreme right instead of 
extreme left
is that in the definition of $\Omega^{(g,m,n)}_p$ there are additional insertions
of $b$-ghosts and other grassmann odd operators besides the operators $A_i$, and we choose
the convention in which they are inserted to the left of the $A_i$'s in a correlation function.
There are precisely $p$ insertions of such grassmann odd operators in the definition on
$\Omega_p$. Therefore
if we want to move the grassmann odd c-numbers to the extreme left, we have to take into account
an extra factor of $(-1)^p$ for each such grassmann odd c-number.} 
and then removing a common
factor involving their product. This is the convention we shall follow throughout.

While we shall not need the explicit form of $\Omega^{(g,m,n)}_p$ 
for our analysis, the following
information will be useful. The grassmann parity of 
$\Omega^{(g,m,n)}_p(K_1,\cdots K_m,L_1,\cdots, L_n)$ is $(-1)^p$ if the states
$\{K_i\}$ and $\{L_j\}$ are all grassmann  even. This arises from the 
$p$ number of additional grassmann odd operators 
inserted in the correlator that defines
$\Omega_p$. These are inserted on the
{\em left} of all the states $\{K_i\}$ and $\{L_j\}$. For even states it does not make any difference,
but for example this is needed for the correct interpretation of \refb{euse} where the
states $Q_BK_i$ and $Q_BL_j$ are grassmann odd. 

Given a pair of points $a\in \wt\PP_{g_1,m_1,n_1}$ and $b\in \wt\PP_{g_2,m_2,n_2}$, we define
by $\{a,b\}$ a one parameter family of points in $\wt\PP_{g_1+g_2, m_1+m_2-2,n_1+n_2}$ as
follows. Let $w_1$ be the local coordinate at the last $NS$ puncture of the Riemann surface 
$a$ and $\wt w_1$ be the local coordinate at the first NS puncture of the Riemann surface $b$.
Then we construct a one parameter family of Riemann surfaces by making the 
identification\footnote{This differs from the convention used in \cite{1703.06410} 
where we used
$e^{-i\theta}$ instead of $e^{i\theta}$. Our $\theta$ parameter is related to that
of \cite{1703.06410} by a minus sign.}
\be \label{etwist}
w_1 \, \wt w_1 = e^{i\theta}, \qquad 0\le\theta < 2\pi\, .
\ee
The resulting Riemann surfaces have genus $g_1+g_2$, $m_1+m_2-2$ NS punctures and 
$n_1+n_2$ R punctures.  The operation \refb{etwist} will be called twist sewing. Similarly
given $a\in \wt\PP_{g_1,m_1,n_1}$ and $b\in \wt\PP_{g_2,m_2,n_2}$, we define by
$\{a;b\}$ a one parameter family of points in $\wt\PP_{g_1+g_2, m_1+m_2,n_1+n_2-2}$ obtained
by twist sewing at R punctures. This means that we identify the local coordinate $z_1$ around
the last R puncture of $a$ and the local coordinate $\wt z_1$ around the first R-puncture of $b$
via
\be \label{etwistR}
z_1 \, \wt z_1 = e^{i\theta}, \qquad 0\le\theta < 2\pi\, .
\ee
This time we also need to insert an extra PCO on the resulting Riemann surface. This is 
done via the insertion of
\be 
\XX_0=\ointop_{|z_1|=1} 
{dz_1\over z_1} \XX(z_1) = \ointop_{|\wt z_1|=1} {d\wt z_1\over \wt z_1} \XX(\wt z_1)\, .
\ee
Therefore $\{a;b\}$ should viewed as a generalized subspace of 
$\wh \PP_{g_1+g_2, m_1+m_2,n_1+n_2-2}$ in the sense described 
below \refb{eaction}
since
we are averaging over many subspaces of $\wh \PP_{g_1+g_2, m_1+m_2,n_1+n_2-2}$,
differing by the location of the PCO.

We are now ready to state the consistency requirement on $\RR_{g,m,n}$. It takes the form
\ben \label{econstr}
\p\RR_{g,m,n} &=& -{1\over 2} \sum_{g_1,g_2\atop g_1+g_2=g} \sum_{m_1,m_2\atop m_1+m_2=m+2}
\sum_{n_1,n_2\atop n_1+n_2=n} {\bf S} [ \{\RR_{g_1,m_1,n_1}, \RR_{g_2,m_2,n_2}\}]
\nonumber \\
&& -{1\over 2} \sum_{g_1,g_2\atop g_1+g_2=g} \sum_{m_1,m_2\atop m_1+m_2=m}
\sum_{n_1,n_2\atop n_1+n_2=n+2} {\bf S} [ \{\RR_{g_1,m_1,n_1} ; \RR_{g_2,m_2,n_2}\}]\, ,
\een
where for example $\{\RR_{g_1,m_1,n_1}, \RR_{g_2,m_2,n_2}\}$ is the result of twist sewing every
Riemann surface of $\RR_{g_1,m_1,n_1}$ with every Riemann surface of $\RR_{g_2,m_2,n_2}$.
${\bf S}$ denotes sum over all inequivalent permutation of the external punctures so that the right
hand side becomes symmetric under the exchange of any pair of NS punctures and any pair of
R punctures. $\p\RR_{g,m,n}$ denotes the boundary of $\RR_{g,m,n}$, excluding the boundaries
associated with non-separating type degeneration of Riemann surfaces. One important feature of the
subspaces $\RR_{g,m,n}$ is that they do not include separating type 
degenerations.\footnote{Possible choices of local coordinate systems satisfying these
requirements can be found in \cite{9206084,1703.10563}.}

The amplitudes of string field theory are given by sum of tree level Feynman diagrams computed from the
1PI effective action \refb{eaction}. For any amplitude with external states 
$K_1,\cdots K_m$,
$L_1,\cdots L_n$ 
there is one diagram that has a single vertex and
no internal propagator -- its contribution is given by $\{K_1\cdots K_m\, L_1\cdots L_n\}$
(up to factor of $i$).  This involves integration over the section segments $\RR_{g,m,n}$ 
which cover only part
of the moduli space $\MM_{g,m,n}$. The contributions from other Feynman diagrams, involving one
or more internal propagators, are given by integration over other section segments. 
Eq.\refb{econstr} guarantees that together these section segments constitute sections of 
$\wt\PP_{g,m,n}$  over the full moduli space $\MM_{g,m,n}$. In particular the regions of the
moduli space containing separating type degenerations arise from Feynman diagrams with one or more propagators.

We end this review by describing the structure of classical equations of motion derived from the action --
this encodes information on the tree level amplitude computed from \refb{eaction} which is turn gives the
loop amplitudes of string theory. The classical equations of motion take the form
\ben \label{eeom}
&& Q_B(|\Psi\rangle - \GG|\wt\Psi\rangle) = 0, \nonumber \\
&& Q_B|\wt\Psi\rangle + \sum_{N\ge 0} {1\over N!} [\Psi^N] = 0\, ,
\een
where for $A_1,\cdots A_N\in \wh\HH$, we define $[A_1\cdots A_N]\in\wt\HH$ via
\be \label{egrconf}
\langle \phi| c_0^-|[A_1\cdots A_N]\rangle = \{\phi\, A_1\cdots A_N\}\, , \quad \forall \, |\phi\rangle
\in \wh\HH\, .
\ee
Multiplying the second equation in \refb{eeom} by $\GG$ and adding it to the first equation we get
\be\label{einteom}
Q_B|\Psi\rangle +  \sum_{N} {1\over N!} \GG\, [\Psi^N] = 0\, .
\ee
This can be regarded as the equation of motion of the interacting field $|\Psi\rangle$. For given 
$|\Psi\rangle$ satisfying \refb{einteom}, we can find $|\wt\Psi\rangle$ by solving the second equation
in \refb{eeom}. Given a particular solution to this equation, 
the only left over freedom in the choice of $|\wt\Psi\rangle$ is the addition of a 
BRST invariant state to the solution. Therefore degrees of freedom associated with $|\wt\Psi\rangle$
describe non-interacting free fields. This can also be  seen from the analysis of the Feynman rules of the
theory\cite{1703.06410}.

Note that in \refb{einteom} we have dropped the summation range over $N$. 
In carrying out our manipulations we shall often not specify the ranges in various sums
with the understanding that the sum ranges over the full set for which the summand does
not vanish.

\subsection{Some new notations} \label{snewnot}

So far we have introduced interaction vertices $\{\cdots\}$  where each external state is an 
off-shell string
field represented by a state in $\wh \HH$. 
However for our analysis we shall also need more general interaction vertices where not all
external states are in $\wh\HH$. 
For this we introduce the space $\wt\PP_{g,m,n;r,s}$ whose base
is the moduli space of genus $g$ Riemann surface with $m$ NS and
$n$ R punctures  in $\wh\HH$, and $r$ NS and $s$ R punctures in $\wt\HH$. The fiber
of $\wt\PP_{g,m,n;r,s}$ contains information on the local coordinates at all the
$m+n+r+s$ punctures and the locations of the $2g-2+m+n/2+r+3s/2$ PCO's
that are needed to get a non-vanishing correlation function.
We can introduce the $p$ forms $\Omega^{(g,m,n;r,s)}_p$ on these spaces by simple
generalization of the definition of $\Omega^{(g,m,n)}_p$ reviewed in \cite{1703.06410}. This
$p$-form satisfies the analog of \refb{euse}:
\ben \label{euse1}
&& \Omega^{(g,m,n;r,s)}_p(Q_B \, K_1,\cdots K_m,L_1,\cdots, L_n;\wt K_1,\cdots \wt K_r,
\wt L_1, \cdots \wt L_s)
+\cdots \nonumber \\ &&
+ \, \Omega^{(g,m,n;r,s)}_p(K_1,\cdots K_m,L_1,\cdots, L_n;
\wt K_1,\cdots \wt K_r,
\wt L_1, \cdots Q_B\wt L_s) \nonumber \\
&=& (-1)^p d\Omega^{(g,m,n; r,s)}_{p-1}(K_1,\cdots K_m,L_1,\cdots, L_n;
\wt K_1,\cdots \wt K_r,
\wt L_1, \cdots \wt L_s)\, .
\een
In order to avoid writing too many indices we shall now introduce a short-hand notation
\be 
\QQ_{g,M,R} = \sum_{m+n=M} \sum_{r+s=R} \wt\PP_{g,m,n;r,s}\, ,
\ee
where the sum simply denotes union. Therefore $Q_{g,M,R}$ is the union of all the
spaces $\wt\PP_{g,m,n;r,s}$ with $m+n=M$ and $r+s=R$. Consequently we also define
$\omega_p^{(g,M;R)}$ to be a $p$-form on $Q_{g,M,R}$ such that
\be \label{edefom}
\omega_p^{(g,M;R)}= (g_s)^{2g} \, 
\Omega_p^{(g,m,n;r,s)} \quad \hbox{when restricted to 
$\wt\PP_{g,m,n;r,s}$}\, .
\ee

\def\dim{{\rm dim}}

\def\rstar{{\rightarrow\atop *}}

\def\lstar{{\leftarrow\atop *}}

\def\rstar{\hskip 6pt * \hskip -12pt \rightarrow}

\def\lstar{\hskip 8pt * \hskip -14pt \leftarrow}

Now let us suppose that $\AAA_{g_1,M_1,N_1}$ and $\BB_{g_2,M_2,N_2}$ 
are 
subspaces of $\QQ_{g_1,M_1,N_1}$ and $\QQ_{g_2,M_2,N_2}$ respectively. 
We define
\be \label{estarint}
\AAA_{g_1,M_1,N_1} * \BB_{g_2,M_2,N_2}\, ,
\ee
as the result of twist sewing $\AAA_{g_1,M_1,N_1}$ and $\BB_{g_2,M_2,N_2}$ similar to the
manner in which we defined it 
in the paragraphs
containing \refb{etwist}-\refb{econstr}, but with some difference:
\begin{enumerate}

\item The sewing is done by picking one of the $M_1$ punctures of $\AAA_{g_1,M_1, N_1}$ 
carrying states in $\wh \HH$ and one of the $M_2$ punctures of $\BB_{g_2,M_2, N_2}$ 
carrying states in $\wh \HH$. Therefore the sewed surfaces form a subspace of 
$\QQ_{g_1+g_2, M_1+M_2-2, N_1+N_2}$.
Since $\AAA$ and $\BB$ are
not necessarily 
symmetric under the exchange of all the punctures, we need to specify which of the
punctures of $\AAA$ and $\BB$ are sewed to each other. In what follows we shall
see that some of the subspaces $\AAA$ and/or $\BB$ carry special punctures where
we always insert a marginal operator. The special punctures are never sewed.
We shall follow the convention that 
leaving aside the special puncture,  
in $\AAA*\BB$ we sew the last $\wh \HH$ puncture  in 
$\AAA$ and first $\wh\HH$ puncture in $\BB$. Most subspaces of interest will be
symmetric under the exchange of all the punctures other than the special puncture; so
it will not make a difference which puncture we use for sewing.

\item Since 
$\AAA$ and $\BB$ are not necessarily even dimensional subspaces, we need to specify the
orientation of the space after gluing. We use the convention that the volume form on the
sewed space is given by the wedge product of the volume form on $\AAA$, $d\theta$ where
$\theta$ is the sewing parameter and the volume form on $\BB$ in this
order. This gives
\be \label{eboundarystar}
\p (\AAA * \BB) = (\p\AAA) * \BB + (-1)^{\dim \AAA + 1} \AAA * \p\BB\, .
\ee
\end{enumerate}

The symbol $*$ describes the effect of sewing two punctures, each in $\wh\HH$. 
As described below \refb{etwistR},  it involves insertion of the operator $\GG$.
We shall
now introduce two new symbols -- $\AAA \rstar \BB$ that sews the last puncture in
$\wh\HH$ from $\AAA$ to the first puncture in $\wt\HH$ in $\BB$ and
$\AAA \lstar \BB$ that sews the last puncture in
$\wt\HH$ from $\AAA$ to the first puncture in $\wh\HH$ in $\BB$, without any insertion
of the operator $\GG$. The quick way to remember the notation is that the arrow points
towards the subspace that contributes the puncture in $\wt\HH$.  Both $\rstar$ and $\lstar$
satisfy identities similar to that in \refb{eboundarystar}.

Given a 
$p$-dimensional subspace $\AAA_{g,M,R}$ of $\QQ_{g,M,R}$, we shall define
\be \label{eshort}
\AAA_{g,M,R}(A_1,\cdots A_M; \wt A_1,\cdots \wt A_R)
\equiv \int_{ \AAA_{g,M,R}} \omega^{(g,M;R)}_p (A_1,\cdots A_M; \wt A_1,\cdots \wt A_R)
\quad \hbox{for} \quad A_i\in \wh\HH, \ \wt A_i\in \wt\HH\, .
\ee
Furthermore if the $k$ of the states $A_1,\cdots A_M$ are equal to some state 
$A$, then we shall express the argument as $A^k$ instead of $k$ copies of $A$.
Similar notation is follows for states in $\wt\HH$. 
In this notation \refb{euse1} can be stated as:
\ben \label{eipart}
&& \AAA_{g,M,R}(Q_B A_1, \cdots A_M; \wt A_1,\cdots \wt A_R) + \cdots
\AAA_{g,M,R}(A_1, \cdots A_M; \wt A_1,\cdots Q_B \wt A_R) \nonumber \\
&=& (-1)^{\hbox{dim}(\AAA_{g,M,R})} \, \p\AAA_{g,M,R}(A_1, \cdots A_M; \wt
A_1,\cdots \wt A_R)
\, .
\een

Let us now consider the quantity 
\be  \label{express1}
\AAA_{g_1,M_1,N_1} * 
\BB_{g_2,M_2,N_2}(A_1,\cdots A_{M_1-1}; \wt A_1,\cdots \wt A_{N_1}|
B_1,\cdots B_{M_2-1};
 \wt B_1,\cdots \wt B_{N_2})\, .
\ee
In writing the arguments on the right hand side we have used the convention that all the states 
inserted into the punctures of $\AAA$ are written first, followed by the states that are
inserted at the punctures of $\BB$, and within each group, the states that are
in $\wh\HH$ are
written first followed by the states in $\wt\HH$.
We shall furthermore assume that $A_i$'s, $B_i$'s, $\wt A_i$'s and
$\wt B_i$'s have been made even, possibly via multiplication by a grassmann odd
c-number. In that case we can write down an expression for \refb{express1} in terms
of the corresponding quantities for component Riemann surfaces as follows. Let $|\vp_r\rangle$ 
and $|\vp^c_r\rangle$ be a conjugate pair of basis states in $\wh\HH\oplus \wt\HH$ satisfying
\be\label{einner}
\langle \vp_r^c| c_0^- | \vp_s\rangle = \delta_{rs} = \langle \vp_s| c_0^- | \vp_r^c\rangle, \quad 
\sum_s |\vp_s\rangle \langle \vp_s^c| = b_0^- = \sum_s |\vp_s^c\rangle \langle \vp_s|\, .
\ee
It is easy to check that $\vp_r$ and $\vp_r^c$ have opposite grassmann parities. 
We shall
denote by $(-1)^{n_r}$ the grassmann parity of $\vp_r$ with $n_r$ taking values 0 or 1.
Then using the completeness relation given in the last equation of \refb{einner} we have
\ben \label{esewing}
&& \AAA_{g_1,M_1,N_1} * \BB_{g_2,M_2,N_2}(A_1,\cdots A_{M_1-1}; \wt A_1,\cdots \wt A_{N_1}|
B_1,\cdots B_{M_2-1};
\wt B_1,\cdots \wt B_{N_2})\nonumber \\
&=& \sum_{r,s}
\AAA_{g_1,M_1,N_1}(A_1,\cdots A_{M_1-1}, \vp_r; \wt A_1,\cdots \wt A_{N_1})\
\BB_{g_2,M_2,N_2}(\vp_s, B_1,\cdots B_{M_2-1}; \wt B_1,\cdots \wt B_{N_2})
\nonumber \\ && \quad \times \, 
\langle \vp_r^c | c_0^- \GG | \vp_s^c\rangle \times (-1)^{n_s\, {\rm dim}\, \BB}\, ,
\een
where the last factor arises from having to move $\vp_s$ inside the argument of 
$\BB_{g_2,M_2,N_2}$ to the extreme left 
through the $\dim\, \BB\equiv \dim\, \BB_{g_2,M_2,N_2}$ 
number of anti-commuting ghost insertions
present in the correlation function that defines $\BB_{g_2,M_2,N_2}$. Similarly  we have
\ben \label{esewing1}
&& \AAA_{g_1,M_1,N_1} \rstar \BB_{g_2,M_2,N_2}(A_1,\cdots A_{M_1-1};
\wt A_1,\cdots \wt A_{N_1}| B_1,\cdots B_{M_2};
\wt B_1,\cdots \wt B_{N_2-1})\nonumber \\
&=& \sum_{r,s}
\AAA_{g_1,M_1,N_1}(A_1,\cdots A_{M_1-1}, \vp_r; \wt A_1,\cdots \wt A_{N_1})
\ \BB_{g_2,M_2,N_2}(B_1,\cdots   B_{M_2}; \vp_s, \wt B_1,\cdots \wt B_{N_2-1})
\nonumber \\ && \quad \times \, 
\langle \vp_r^c | c_0^- | \vp_s^c\rangle \times (-1)^{n_s\, {\rm dim}\, \BB}\, ,
\een
and 
\ben \label{esewing2}
&& \AAA_{g_1,M_1,N_1} \lstar \BB_{g_2,M_2,N_2}(A_1,\cdots A_{M_1};
\wt A_1,\cdots \wt A_{N_1-1}| B_1,\cdots B_{M_2-1};
\wt B_1,\cdots \wt B_{N_2})\nonumber \\
&=& \sum_{r,s}
\AAA_{g_1,M_1,N_1}(A_1,\cdots A_{M_1}; \wt A_1,\cdots \wt A_{N_1-1}, \vp_r)\, 
\BB_{g_2,M_2,N_2}( \vp_s, B_1,\cdots B_{M_2-1}; \wt B_1,\cdots \wt B_{N_2})
\nonumber \\ && \quad \times \, 
\langle \vp_r^c | c_0^- | \vp_s^c\rangle \times (-1)^{n_s\, {\rm dim}\, \BB}\, .
\een
Using these relations and ghost number conservation rules one can
show that if $\AAA_{g_1,M_1,N_1}$ and $\BB_{g_2,M_2,N_2}$ are symmetric under the exchange
of the punctures in $\wh \HH$ and the exchange of the punctures in $\wt\HH$, then,
\ben \label{eexch1}
&& \AAA_{g_1,M_1,N_1} * \BB_{g_2,M_2,N_2}(A_1,\cdots A_{M_1-1};
\wt A_1,\cdots \wt A_{N_1}| B_1,\cdots B_{M_2-1};
\wt B_1,\cdots \wt B_{N_2}) \nonumber \\
&=& \BB_{g_2,M_2,N_2} * \AAA_{g_1,M_1,N_1} (B_1,\cdots B_{M_2-1};
\wt B_1,\cdots \wt B_{N_2} | A_1,\cdots A_{M_1-1};
\wt A_1,\cdots \wt A_{N_1})\nonumber \\
&& \qquad \qquad \times \ (-1)^{\dim \AAA \, \dim\BB + \dim \AAA + \dim\BB}\, ,
\een
for even states $\{A_i, B_i, \wt A_i, \wt B_i\}$. Similarly we have 
\ben \label{eexch2}
&& \AAA_{g_1,M_1,N_1} \lstar \BB_{g_2,M_2,N_2}(A_1,\cdots A_{M_1};
\wt A_1,\cdots \wt A_{N_1-1}| B_1,\cdots B_{M_2-1};
\wt B_1,\cdots \wt B_{N_2}) \nonumber \\
&=& \BB_{g_2,M_2,N_2} \rstar \AAA_{g_1,M_1,N_1} (B_1,\cdots B_{M_2-1};
\wt B_1,\cdots \wt B_{N_2}|  A_1,\cdots A_{M_1};
\wt A_1,\cdots \wt A_{N_1-1})\nonumber \\
&& \qquad \qquad \times \ (-1)^{\dim \AAA \, \dim\BB + \dim \AAA + \dim\BB}\, .
\een
These relations can also be derived using the definition of the volume form on 
$\AAA *\BB$ (and similar definitions for $\AAA\lstar\BB$ and
$\AAA\rstar\BB$) given above \refb{eboundarystar}.

We shall streamline the notations even further by defining 
the formal sum
\be
\QQ_{M,N}\equiv \sum_{g\ge 0} \QQ_{g,M,N}\, .
\ee
Similarly given a family of subspaces $\AAA_{g.M,N}$ of $\QQ_{g,M,N}$ we shall
define 
\be\label{esimpli}
\AAA_{M,N} \equiv \sum_{g\ge 0} \AAA_{g,M,N}\, ,
\ee
so that $\AAA_{M,N}\subset \QQ_{M,N}$.
Therefore when we have a symbol with two subscripts it will be understood that we have
summed over $g$.

Let us now rewrite some of the old identities in the new notation. 
If we define
\be
\VV_{g,M,0} \equiv \sum_{m,n\atop m+n=M} \, \RR_{g,m,n}\, ,
\ee
describing a subspace of $\QQ_{g,M,0}$, and
\be 
\VV_{M,0} = \sum_{g\ge 0} \VV_{g,M,0}\, ,
\ee
then we have
\be \label{estring}
(g_s)^{2g}\, \{A_1\cdots A_M\}_g = \VV_{g,M,0}(A_1,\cdots A_M), \qquad 
\{A_1\cdots A_M\} = \VV_{M,0}(A_1,\cdots A_M)\, .
\ee
In this notation, eq.\refb{econstr} may be expressed as
\be \label{evvcompact}
\p\VV_{M,0} (\Phi^M)= -{1\over 2} \sum_{M_1, M_2\atop
M_1+M_2=M} {M!\over M_1! M_2!}\,  \VV_{M_1+1,0}* \VV_{M_2+1,0}(\Phi^{M_1}|\Phi^{M_2})\, ,
\ee 
where the combinatorial factor arises from symmetrization operation ${\bf S}$ in
\refb{econstr}.
To recover \refb{econstr} from here we need to take $\Phi=A_1+\cdots +A_M$ and pick the
coefficient of $A_1\cdots A_M$ from both sides. 
Using \refb{eipart} we can also express \refb{evvcompact} as
\be\label{enewrr}
M \, \VV_{M,0}(Q_B\Phi,\Phi^{M-1})
= -{1\over 2}\, \sum_{M_1, M_2\atop
M_1+M_2=M} {M!\over M_1! M_2!}\,  \VV_{M_1+1,0}* \VV_{M_2+1,0}(\Phi^{M_1}|\Phi^{M_2})\, .
\ee
This is the form of the relation that is directly used to prove the invariance of the 1PI effective action
\refb{eaction} under the infinitesimal gauge transformation:
\be 
\delta |\wt\Psi\rangle = Q_B|\wt\Lambda\rangle + \sum_M {1\over M!} \, [\Lambda \Psi^M]\,,
\qquad \delta |\Psi\rangle = Q_B|\Lambda\rangle 
+ \sum_M {1\over M!} \,\GG[\Lambda \Psi^M]\, ,
\ee
where $\Lambda\in\wh\HH$ and $\wt\Lambda\in\wt\HH$ are infinitesimal grassmann odd
gauge transformation parameters.

\sectiono{Statement of background independence} \label{s3}

In this section we shall discuss the precise statement of background independence of
superstring field theory that we shall attempt to prove.

\subsection{Marginal deformation of the superconformal field theory}

Let us suppose that the matter sector of the world-sheet SCFT admits
a supersymmetry preserving marginal deformation. 
In the heterotic world-sheet theory this implies the existence of a
dimension $(1,1/2)$ superconformal primary operator $\OO$ of the matter 
theory.\footnote{We use the convention that left-movers are anti-holomorphic and
right-movers are holomorphic fields on the world-sheet.}
We also have an associated dimension $(1,1)$ matter primary $\wt\OO$
defined through the operator product expansion
\ben \label{eope1}
&& T_F(z) \, \OO(w,\bar w) \simeq (z-w)^{-1} \, \wt\OO(w,\bar w) + \hbox{less singular terms}\, ,
\nonumber \\
&& T_F(z) \, \wt\OO(w,\bar w) \simeq {1\over 4} (z-w)^{-2} \OO(w,\bar w) + {1\over 4} (z-w)^{-1} \p\OO(w, \bar w)\
+ \hbox{less singular terms}\, . \nonumber \\
\een
$\OO$ and $\wt\OO$ may be considered as the lower and the upper components of a
superfield.
We can now consider a neighboring SCFT
that is related to the original theory by
marginal deformation by the operator $\wt\OO$.
Without loss of
generality we shall restrict our discussion to first order in the deformation parameter since we can build
a finite deformation by successive application of infinitesimal deformations.
In this case
the correlation functions of the new theory on any Riemann surface can be computed 
in terms of correlation functions
of the original theory on the same Riemann surface by making an additional insertion of the marginal operator
and integrating over the location of its insertion point.
More precisely we deform every correlation function by inserting into the correlation
function an operator
\be \label{epert1a}
 - {\lambda\over 2\pi i} \,\int dz\wedge d\bar z \, \wt \OO(z,\bar z)\, ,
\ee
for some infinitesimal operator $\lambda$.
However this
apparently suffers from
divergences when the locations of the marginal operator approach the locations of the other operators
inserted in the correlation function.
It also suffers from an ambiguity since in principle the states of the undeformed theory and the 
states of the deformed theory belong to different Hilbert spaces and there is no canonical isomorphism
between these Hilbert spaces. Therefore by state operator correspondence, there is also no 
canonical isomorphism 
between the local operators in the two theories, and there is no absolute notion of comparing 
correlation functions in the two theories.

It turns out that both problems can be resolved by introducing the notion of a
connection on the
Hilbert space of the family of SCFT's related by marginal deformation. This 
establishes a (non-canonical) isomorphism between the Hilbert spaces
of the original theory and the deformed theory. 
Once such an isomorphism is established, it becomes meaningful to express the correlation function
of one theory in terms of the correlation function of the other theory. It turns out that when we choose a
non-singular connection it also eliminates the divergence problems. We shall now state the
result for the deformed correlation function 
for a particular choice of connection\cite{as,nelson-c,ranga1,ranga2}. 
In that case the correlation function of a set of operators 
$A_1,\cdots A_N$ on a Riemann surface $\Sigma$, inserted using local coordinates 
$w_1,\cdots w_N$, changes by
\be\label{edeform}
\delta  \langle \, A_1\cdots A_N \rangle_\Sigma
= -{\lambda \over 2\pi i}
\int_{\Sigma - \cup_i D_i}\,  dz\wedge d\bar z \, \langle \, \wt\OO(z,\bar z) \, A_1\cdots A_N \rangle_\Sigma
\ee
where $\Sigma - \cup_i D_i$ denotes the whole Riemann surface sans unit disks $|w_i|\le 1$ 
around each puncture. Since in \refb{edeform} the location $z$ of the additional operator $\wt \OO$
never approaches any of the punctures, there is no divergence of the type mentioned earlier. The
fact that we have to exclude disks of unit radius from around each puncture is dictated by the
special choice of connection we have taken. Excluding disks of radius $a$ for any other number
$a$ will correspond to a different
choice of connection. $a\to 0$ limit corresponds to the original prescription of integrating over the 
whole Riemann surface. In this limit we get back the divergences mentioned earlier but this can now be
attributed to the fact that the corresponding choice of connection is singular. The 
particular choice of connection described in \refb{edeform} 
preserves the BPZ inner product.

Since the PCO's are typically inserted on $\Sigma-\cup_i D_i$, we can get divergences when the location
$z$ of $\wt\OO$ approaches the location $w$ of the PCO. However since the divergent piece is 
proportional to $(z-w)^{-2}$, we can perform the integral by restricting the
$z$ integral to $|z-w|>\eps$ for some small number $\eps$ and then taking the $\eps\to 0$ limit.
This gives a finite integral and furthermore the result of integration is not affected under a non-singular
change of coordinates i.e. we get the same result by performing the integral over 
the region $|f(z)-f(w)|>\eps$ and then taking the $\eps\to 0$ limit. For this reason we shall not worry
about collision of $\wt\OO$ with the PCO's.

For the particular choice of connection that excludes disks of unit radius, 
we can calculate the 
deformations of the standard SCFT operators like the Virasoro 
generators and the modes of the
supercurrent. The modes of the ghost fields remain unchanged since the deformation is
only in the matter sector. The Virasoro generators $L_n$ and $\bar L_n$ change 
by\cite{as,nelson-c,ranga1,ranga2}
\be \label{eldeform}
\delta L_n =\lambda\, \ointop_{|z|=1} d\bar z \, z^{n+1} \, \wt\OO(z,\bar z), \qquad 
\delta \bar L_n = \lambda\, \ointop_{|z|=1} d z \, \bar z^{n+1} \, \wt\OO(z,\bar z)\, ,
\ee
where the integral is performed over a circle at $|z|=1$. 
$\ointop$ includes appropriate factors of $\pm (2\pi i)^{-1}$ so that $\ointop dz/z=1$,
$\ointop d\bar z /\bar z =1$.
The modes $G_n$ of the matter
supercurrent change by\cite{nelson-c}\footnote{The sign and normalization of the right hand side
of \refb{edgn} differs from that in \cite{nelson-c} due to a difference in convention. 
The correct normalization can be
derived  using the procedure described in \cite{as} using \refb{eope1}.}
\be\label{edgn}
\delta G_n= {1\over 4}\lambda\,  \ointop_{|z|=1} d\bar z \, z^{n+1/2} \, \OO(z,\bar z)\, .
\ee
As a result the BRST operator $Q_B$ and the zero mode of the PCO $\XX_0$ 
change by
\be \label{e3.6}
\delta Q_B = \lambda\, \ointop_{|z|=1} d\bar z \, c(z) \, \wt\OO(z,\bar z) +  \lambda\, 
\ointop_{|z|=1} d z \, \bar c(\bar z) \, \wt\OO(z,\bar z) +{1\over 4} \lambda\,  \ointop_{|z|=1} d\bar z \,
\, \gamma(z) \, \OO(z,\bar z)\, ,
\ee
and
\be\label{e3.7}
\delta \XX_0 ={1\over 4}\, 
\lambda\,  \ointop_{|z|=1} d\bar z \, z^{-1} \, e^{\phi}(z) \, \OO(z,\bar z)\, .
\ee
It is straightforward (although somewhat tedious) to verify that 
\be\label{ennxx}
\{Q_B, \delta Q_B\} = \OO(\lambda^2), \qquad [Q_B, \delta\XX_0] + [\delta Q_B, \XX_0]
=\OO(\lambda^2)\, ,
\ee
so that $Q_B+\delta Q_B$ is nilpotent and commutes with $\XX_0+\delta
\XX_0$ to first order in the perturbation parameter $\lambda$. 
The following argument also shows that 
the connection
preserves the projection into the $L_0^-=0$ states, i.e.  $\delta L_0 - \delta\bar L_0$
vanishes while acting on a state with $L_0=\bar L_0$. Let us suppose that $\phi$
is a vertex operator with $L_0=\bar L_0=h$. Then we have the operator product expansion
\be
\wt\OO(z,\bar z) \, \phi(0) =\sum_{p,q} \bar z^{p-1} \, z^{q-1} \, \phi_{p,q}(0)\, ,
\ee
where $\phi_{p,q}$ have dimension $(p+h, q+h)$. Therefore
\be
(\delta L_0 - \delta \bar L_0)|\phi\rangle 
=\lambda\,  \ointop_{|z|=1} d\bar z \, z\, \sum_{p,q} \bar z^{p-1} \, 
z^{q-1} \, |\phi_{p,q}\rangle 
- \lambda\, \ointop_{|z|=1} d z \,  \bar z \, \sum_{p,q} \bar 
z^{p-1} \, z^{q-1} \, |\phi_{p,q}\rangle 
\ee
Since the contour integrals are performed over circles of unit radii around the origin,
each of these terms vanish unless $p=q$. On the other hand  for $p=q$ the two terms 
gives identical result $|\phi_{p,q}\rangle$ and cancel. This shows that 
$\delta L_0 - \delta\bar L_0$
vanishes while acting on a state with $L_0=\bar L_0$. Since the $b_n$'s are not deformed
at all, the $b_0-\bar b_0=0$ condition is also preserved by this connection.

We can now consider the string field theory action $S'(\Pi,\wt\Pi)$ formulated in the new
background. It is given by
\be \label{e1.28}
S'(\Pi,\wt\Pi) =S(\Pi,\wt\Pi) + \delta S(\Pi,\wt\Pi)\, ,
\ee
where
\be \label{e1.29}
\delta S = g_s^{-2} \left[ -{1\over 2} \langle \wt\Pi| c_0^- (\delta \GG \, Q_B 
+  \GG \, \delta Q_B ) |\wt\Pi\rangle
+ \langle \wt\Pi| c_0^-  \delta Q_B |\Pi\rangle + \sum_N {1\over N!} \delta\{\Pi^N\}\right]\, .
\ee
Here $\delta\{A_1\cdots A_N\}$ denotes the change in $\{A_1\cdots A_N\}$ due to the
change in the correlation function given in \refb{edeform}. 

The equations of motion for the interacting part of the theory is obtained by replacing 
the kinetic and interaction terms in \refb{einteom} by their counterpart in the new background. 
This gives
\be \label{eomdeformed}
Q_B|\Pi\rangle + \sum_{N} {1\over N!} \, \GG\, [\Pi^N] + \delta Q_B\, |\Pi\rangle 
+ \sum_{N} {1\over N!} \, \delta\, \GG\, [\Pi^N] + \sum_{N} {1\over N!} \, \GG\, 
\delta\, [\Pi^N] =0 \, ,
\ee
where $\delta\, [\Pi^N]$ is defined via the equation
\be 
\langle A| c_0^- |\delta\, [\Pi^N]\rangle = \delta \, \{A\, \Pi^n\}
\ee
for any  state $A$.

\subsection{Superstring field theory around shifted background}

Given the superconformal primary operator $\OO(z,\bar z)$ of dimension (1,1/2),
we can construct a BRST invariant operator $\bar c c e^{-\phi}\OO$. Since the genus 
zero contribution 
$\{A_1\cdots A_N\}_0$ vanishes for $N\le 3$, the interaction terms begin at cubic order.
Therefore
\be \label{edefpsi0}
|\Psi\rangle =|\wt\Psi\rangle = \lambda\, |\PSZ\rangle, \qquad
|\PSZ\rangle \equiv  \bar c_1\, c_1\, e^{-\phi}(0) |\OO\rangle\, ,
\ee
gives a solution to the {\it classical} equations of motion \refb{eeom} of the 
undeformed theory to order $\lambda$. 
Note that since the solution carries picture number $-1$ and therefore describes
an NS sector state, $\GG=1$ in this sector and we can take 
$\wt\Psi=\Psi$.\footnote{Since the equations of motion allow us to shift $\wt\Psi$ by a BRST
invariant state without affecting $\Psi$, we could set $\wt\Psi=0$.}
We shall not demand that $\lambda|\PSZ\rangle$ remains a solution to the 
quantum corrected equations of motion, just as we do not assume that $|\Pi\rangle=0$
is a solution to the deformed equation of motion \refb{eomdeformed} after including
the quantum effects in the 1PI effective action.

We now
define shifted fields
\be 
|\Phi\rangle = |\Psi\rangle -\lambda\,  |\PSZ\rangle, \quad |\wt\Phi\rangle = |\wt\Psi\rangle
-\lambda\, |\PSZ\rangle\, ,
\ee
and expand the action in a power series expansion in $|\Phi\rangle$. 
The action
takes the form
\be 
S(\Psi,\wt\Psi)= S(\PSZ, \PSZ) + S''(\Phi, \wt\Phi)\, ,
\ee
where 
\be \label{edefact}
S''(\Phi, \wt\Phi)= g_s^{-2} \left[ -{1\over 2} \langle \wt\Phi| c_0^- \GG \, Q_B |\wt\Phi\rangle
+ \langle \wt\Phi| c_0^-  Q_B |\Phi\rangle + 
\sum_N {1\over N!} \{\Phi^N\} + \lambda \, \sum_N {1\over N!} \{\PSZ\, \Phi^N\} \right]
+\OO(\lambda^2)\, .
\ee
The equations of motion derived from
\refb{edefact} takes the form
\ben \label{eeomsh}
&& Q_B(|\Phi\rangle- \GG|\wt\Phi\rangle)=0 \nonumber \\ &&
Q_B|\wt\Phi\rangle + \sum_N {1\over N!} [\Phi^N] + \lambda \, \sum_N {1\over N!} [\PSZ\,
\Phi^N]
=0\, .
\een
Multiplying the second equation in \refb{eeomsh} by $\GG$ and adding it to the first equation
we get the equation of motion for the interacting field $|\Phi\rangle$:
\be \label{eeomsh2}
Q_B |\Phi\rangle + \sum_N {1\over N!} \, \GG \left([\Phi^N]
+\lambda [\PSZ \, \Phi^N]\right) =0\, .
\ee
For given $|\Phi\rangle$ satisfying \refb{eeomsh2}, the second equation of \refb{eeomsh}
determines $|\wt\Phi\rangle$ up to addition of BRST invariant states. Therefore as before,
the degrees of freedom of $\wt\Phi$ are free fields, and \refb{eeomsh2} contains full
information about the S-matrix of the interacting part of the theory.

\subsection{Statement of the problem}

We now have two potential descriptions of quantum corrected interacting field equations of 
superstring field theory in a background related
to the original background via marginal deformation.  The first one, described by 
\refb{eomdeformed}, 
uses the formulation of superstring field theory around the deformed
world-sheet SCFT. The second one, described by \refb{eeomsh2} is the equation of motion
of 
superstring field theory formulated around the original background, but expanded around
a new solution to the equations of motion 
representing the marginal deformation. The statement of
background independence is that these two sets of equations are equivalent 
(to order $\lambda$).
Therefore there must be a field redefinition relating $\Pi$ to $\Phi$
that makes the first set of equations of motion into  linear combinations of
the second set (with possibly field dependent coefficients).\footnote{The advantage 
of working with the equations of motion derived from the 1PI effective
action is that we do not need to worry about the change in the integration measure over
fields under a change in background, since the effect of integration measure is included
in the definition of the 1PI effective action. This avoids some of the complications encountered
in \cite{back2}.} Since the two sets of equations of motion differ by order $\lambda$, 
we can assume that the (field dependent) matrix relating the two sets of equations
differ from the identity matrix by terms of order $\lambda$. Denoting the original
equations of motion by $|E_0\rangle=0$, and the equations \refb{eomdeformed} and
\refb{eeomsh2} by $|E_0\rangle+\lambda|E_1\rangle=0$ and 
$|E_0\rangle+\lambda|E_2\rangle=0$ respectively, we can state the requirement as
\be 
|E_0\rangle+\lambda|E_1\rangle = (1 + \lambda \, M) (|E_0\rangle+\lambda|E_2\rangle )
\ee
where $M$ is some linear operator on $\wh \HH$ that could be field dependent. To
order $\lambda$ this gives
\be \label{ereqd}
|E_1\rangle -|E_2\rangle = M \, |E_0\rangle\, .
\ee

We shall look for a field redefinition of the form
\be \label{efredef}
|\Pi\rangle = |\Phi\rangle +  \delta \,  |\Phi\rangle \, ,
\ee
for some state $\delta \, |\Phi\rangle$ of order $\lambda$, and
analyze the difference between the equations
of motion \refb{eomdeformed} and \refb{eeomsh2} to order $\lambda$.
The inner product between an
arbitrary state $\wt A\in \wt\HH$ and  this difference may then be written as
\ben \label{eDelta}
\Delta &\equiv&  \langle \wt A| c_0^- \Bigg( Q_B \,  |\delta \Phi\rangle 
+ \sum_{N} {1\over N!} \GG [\Phi^N \, \delta\Phi] + \delta Q_B |\Phi\rangle
+ \sum_N {1\over N!} \delta \GG \, [\Phi^N] + \sum_N {1\over N!} \GG\, \delta [\Phi^N] \nonumber \\ &&
 - \lambda \, \sum_N {1\over N!} \GG \,
[\PSZ\, \Phi^N]
\Bigg)\, .
\een
For definiteness we shall take $\wt A$ to be grassmann even -- this can always be achieved by multiplying the
state by a grassmann odd c-number if needed. In that case \refb{eDelta} may be written as
\ben \label{eDeltanew}
\Delta &=&  \langle Q_B\wt A| c_0^-  |\delta \Phi\rangle 
+ \sum_{N} {1\over N!} \{ (\GG \wt A) \Phi^N \, \delta\Phi\} + \langle \wt A|c_0^- 
\delta Q_B |\Phi\rangle
+ \sum_N {1\over N!} \{ (\delta \GG \wt A) \, \Phi^N\}\nonumber \\ &&
+ \sum_N {1\over N!} \delta \{
(\GG\, \wt A) \, \Phi^N\} - \lambda \, \sum_N {1\over N!} \{ (\GG \wt A) \,
\PSZ \Phi^N\}
\, .
\een
Eq.\refb{ereqd} now translates to the requirement that 
$\Delta$ vanishes to order $\lambda$ when $|\Phi\rangle$ satisfies the zeroth order 
equation of motion
\be\label{eleading}
Q_B|\Phi\rangle + \sum_N {1\over N!} \, \GG\, [\Phi^N]=0\, .
\ee
Our goal will be to prove the existence of $\delta\Phi$ satisfying this requirement.

One could have asked for more -- demanding that the two actions $S'(\Pi, \wt\Pi)$ and
$S''(\Phi, \wt\Phi)$ are equivalent. However it is easy to see that this cannot hold.
Since the equations of motion of $\Pi, \wt\Pi$ use the deformed BRST operator $Q_B'$
and deformed interaction terms, it follows that the free field degrees of freedom encoded 
in $\wt\Pi$ correspond to states annihilated by $Q_B'$. On the other hand it follows
from \refb{eeomsh} that the free field degrees of freedom encoded in $\wt\Phi$ 
correspond to states annihilated by $Q_B$. Since $Q_B$ and $Q_B'$ have different
cohomologies (in particular the mass spectrum computed from the two operators are
different) there cannot be a field redefinition that relates the two actions. However this
is not necessary for the proof of background dependence -- it is enough to show that
the 1PI equations of motion for the interacting fields are related by field redefinition. This
is what we shall attempt to do.

\sectiono{Geometric interpretation of the kinetic and interaction terms of the deformed theory} \label{sgeo}

In this section we shall 
give geometric interpretation of  $\delta Q_B$, $\delta\GG$ and $\delta\{\Phi^N\}$ by associating to each of them appropriate subspaces of $\QQ_{g,M,N}$. This will be used
in the next section to convert \refb{eDeltanew} into a geometric form.

\subsection{Deformation of the BRST operator}

Let us begin with $\delta Q_B$. 
For this we
introduce a zero dimensional subspace $\VV'_{0,2,1}\subset
\QQ_{0,2,1}$ of the following form. If $z$ denotes the standard coordinate on the 
complex plane used to parametrize a genus 0 surface, then the second puncture in 
$\wh \HH$
and the puncture in
$\wt\HH$ are taken to be at $z=0$ and $z=\infty$ respectively,
with $z$ and $1/z$ as local coordinates
around the two punctures. The first puncture in $\wh\HH$ is inserted at $z=1$. This is
a special puncture where we always insert the vertex operator $\PSZ$  described in
\refb{edefpsi0}\cite{back1,back2}. 
Since this is a dimension zero primary we do not need to specify the
choice of local coordinates at this puncture -- all choices are equivalent. 
Since the total picture number carried by the vertex operators $\PSZ$, and 
the two vertex operators in $\wh \HH$ and $\wt\HH$
is $-3$, we need one PCO insertion. 
This
is inserted at the special puncture at $z=1$, converting 
$\Psi_0$ to a zero picture vertex operator 
\be \label{ezero}
\lim_{w\to z} \XX(w) \PSZ(z) =
c\, \bar c\, \wt\OO - {1\over 4} \, \gamma \, \bar c\, \OO\, ,
\ee
without
generating any singular term.
In this case in the notation of \refb{eshort},
\be\label{eint00}
\VV'_{0,2,1} (\PSZ,  B; \wt A)
= \left\langle \wt A\left| \left[ c\, \bar c\, \wt\OO(1) - {1\over 4} \, \gamma \, 
\bar c\, \OO(1)
\right] \right|B \right\rangle 
\ee
for grassmann even vertex operators $A$ and $\wt B$.
Using the invariance of
states in $\wh\HH$ and
$\wt\HH$ under $L_0^-$ and $b_0^-$, and standard identities for
sphere three point function in
conformal field theory, we can bring this to the form
\ben \label{eint1}
\VV'_{0,2,1} (\PSZ,  B; \wt A)
&=&
\Bigg\langle \wt A\Bigg| c_0^-\,  \Bigg[\ointop_{|z|=1} d\bar z \, c(z) \, \wt\OO(z,\bar z) 
+ \ointop_{|z|=1} d z \, \bar c(\bar z) \, \wt\OO(z,\bar z) 
\nonumber \\ && 
 +{1\over 4}  \ointop_{|z|=1} d\bar z \,
\, \gamma(z) \, \OO(z,\bar z)\Bigg]
\Bigg|B \Bigg\rangle 
\nonumber \\ &=& \lambda^{-1} \, \langle \wt A| c_0^-
\delta Q_B
|B \rangle \, .
\een
To see how this works, let us examine the second term on the right hand side of 
\refb{eint00}. We first write
\be\label{eint01}
\bar c(1) = \sum_n \bar c_n =- {1\over 2} \sum_n (c_n-\bar c_n) + {1\over 2}
(c_n+\bar c_n)\, ,
\ee
and note that since $|\wt A\rangle$ and $|B\rangle$ are annihilated by $b_0^-$,
the only term in \refb{eint01} that contributes to \refb{eint00} is the
$-(c_0-\bar c_0)/2=-c_0^-$ term.
This reduces the second term on the right hand side of \refb{eint00} to
\be 
{1\over 4} \, \langle \wt A| c_0^- \gamma(1) \, 
\OO(1)
|B\rangle \, .
\ee
Denoting by $(h_A, h_A)$ and $(h_B, h_B)$ the conformal weights of
$|\wt A\rangle$ and $|B\rangle$, and recalling that $\gamma\, \OO$ has conformal weight
$(1,0)$, we get
\ben 
\langle \wt A| c_0^- \gamma(z) \, \OO(z,\bar z)
|B\rangle &=& \bar z^{h_A -h_B-1}  z^{h_A - h_B} \langle \wt A| c_0^- \gamma(1) \, 
\OO(1)
|B\rangle =  \bar z^{-1}\, \langle \wt A| c_0^- \gamma(1) \, 
\OO(1)
|B\rangle\, , \nonumber \\ &&  \hbox{at $|z|=1$}\, .
\een
Therefore we get 
\be 
{1\over 4} \ointop_{|z|=1} d\bar z\, \langle \wt A| c_0^- \gamma(z) \, \OO(z,\bar z)
|B\rangle = {1\over 4} \langle \wt A| c_0^- \gamma(1) \, 
\OO(1)
|B\rangle\, .
\ee
This establishes the equality of the second term on the right hand side of 
\refb{eint00} and the term in the second line of \refb{eint1}. Similar
manipulations can be carried out for the other terms\cite{back1,back2}.

From \refb{eint1} we have
\be \label{eddqb}
\langle \wt A| c_0^-
\delta Q_B
|B \rangle =\lambda\, \VV'_{0,2,1} (\PSZ,  B; \wt A) \, .
\ee
Note that since $\VV'_{0,2,1}$ is a zero dimensional subspace, the right hand side of
\refb{eddqb} corresponds to simply evaluating 
$\omega^{(0,2;1)}_0(\PSZ, B;\wt A)$
on a specific point in $\QQ_{0,2,1}$ that represents $\VV'_{0,2,1}$.
Also $\p\VV'_{0,2,1}=0$. When $\wt A$ and $B$
have general grassmann parities $(-1)^{\wt A}$ and $(-1)^B$, 
then this equation will have an additional sign of  $(-1)^{\wt A\, B}$.
This can be seen by multiplying $\wt A$ and $B$ by grassmann odd c-numbers if needed
to make them grassmann even, applying \refb{eddqb} and finally stripping off the
grassmann odd c-numbers by moving them to the extreme right or extreme left.

\subsection{Deformation of the picture changing operator}

Next we turn to $\delta\GG$.
For this we define a one dimensional subspace 
$\VV''_{0,1,2}$ of $\QQ_{0,1,2}$ as follows. 
If any of the punctures in $\wt\HH$ is
NS puncture, we  
declare $\VV''_{0,1,2}$ to be zero, reflecting the fact that
$\delta\GG$ vanishes on NS sector state.
When the two punctures in $\wt\HH$ are Ramond punctures, $\VV''_{0,1,2}$
describes a one dimensional subspace of $\QQ_{0,1,2}$ such that for each
element of $\VV''_{0,1,2}$
the choice of local coordinates at the punctures are fixed
in the same way as for $\VV'_{0,2,1}$ at all the punctures.  In particular
the puncture in $\wh\HH$ is a special puncture at $z=1$
where
$\PSZ$ is inserted,
the first puncture in $\wt\HH$ is at $z=0$ 
with local coordinate
$z$ and the second puncture
in $\wt\HH$ is at $z=\infty$ with local coordinate $1/z$.
Therefore the
different elements  of $\VV''_{0,1,2}$ 
differ only in the choice of PCO locations, making this a `vertical
segment' in the language of \cite{1408.0571,1504.00609}. This one dimensional
vertical segment
interpolates between the following pair of PCO configurations. In both configurations
one PCO is inserted
at the special puncture, converting $\PSZ$ into a zero picture vertex operator given in 
\refb{ezero}.  The second 
PCO is inserted as $\XX_0$ around the puncture at $\infty$ in the initial configuration
and $\XX_0$ around the
puncture at 0 in the final configuration. 
In the correlation function that defines $\VV''_{0,1,2}(\PSZ,\wt A, \wt B)$,
this corresponds to the difference between
two terms -- an insertion of $\xi_0$ around the puncture at $\infty$ 
and the insertion of $\xi_0$ at the
puncture at $0$\cite{1408.0571}.\footnote{Even though $\xi_0$ is not a good operator
in the small Hilbert space in which we are working, the difference between two insertions
of $\xi_0$ is in the small Hilbert space.}
Therefore we have, for grassmann even $\wt A$, $\wt B$:
\be \label{exs1}
\VV''_{0,1,2}(\PSZ, \wt A, \wt B) = \left\langle \wt B\left| 
\left[\xi_0, \left( c\, \bar c\, \wt\OO(1) - {1\over 4} \, 
\gamma \, 
\bar c\,\OO(1)
\right)\right] \right|\wt A\right\rangle
= -{1\over 4} \, \langle \wt B| e^\phi \, \bar c\, \OO (1)|\wt A\rangle\, .
\ee
Following the same logic described below \refb{eint1} we can express this as
\be \label{exs2}
\VV''_{0,1,2}(\PSZ, \wt A, \wt B) =
-{1\over 4} \, \left\langle \wt B\left | c_0^- \ointop_{|z|=1} {d\bar z\over z} 
e^\phi \OO(z,\bar z)\right |\wt A\right\rangle
= - \lambda^{-1} \, \langle \wt B| c_0^- \, \delta\XX_0|\wt A\rangle\, .
\ee
Therefore we may write, for grassmann even states $\wt A,\wt B\in\wt\HH$,
\be\label{egenpar}
\langle \wt B| c_0^- \, \delta\GG|\wt A\rangle = - \lambda\, \VV''_{0,1,2}(\PSZ; \wt A, \wt B)\, .
\ee
When $\wt A$ and $\wt B$ are NS sector states then both sides vanish and the
equation holds identically. When $\wt A$ and $\wt B$ are R sector states then this
equation follows from the equality of \refb{exs1} and \refb{exs2}. When 
$\wt A$, $\wt B$ have general
grassmann parities then this equation will have an additional factor of
$(-1)^{\wt B+\wt B\wt A}$.
This extra sign arises from the fact  that when we convert $\wt A$ and $\wt B$
to grassmann even operators by multiplying them by (possibly) grassmann odd
c-numbers $\zeta_A$ and $\zeta_B$ from the right so that \refb{egenpar} holds,
and 
try to strip off the $\zeta$'s by moving them to the extreme  right on both sides in the 
combination $\zeta_B\zeta_A$,
we get a factor of  $(-1)^{\wt B+\wt B\wt A}$ on the left hand side 
arising from the effect of passing $\zeta_B$ through $c_0^-$ and $\wt A$. An identical
result is obtained by moving the $\zeta$'s to the extreme left.

Since $\VV''_{0,1,2}$ interpolates between two configurations: $\VV'_{0,2,1}$ with an
extra factor of $\GG$ inserted around $\infty$  and $\VV'_{0,2,1}$ with an
extra factor of $\GG$ inserted around 0, we have
\be \label{evvppboundary}
\p\VV''_{0,1,2}(\PSZ; \wt A, \wt B) = \VV'_{0,2,1}(\PSZ, \GG \wt A; \wt B)
- \VV'_{0,2,1}(\PSZ, \GG \wt B; \wt A)\, ,
\ee
for grassmann even
$\wt A$ and $\wt B$.
Another useful property of $\VV''_{0,1,2}(\PSZ; \wt A, \wt B)$ is that it is antisymmetric under
$\wt A\leftrightarrow\wt B$ for grassmann even
$\wt A$ and $\wt B$, 
\be \label{evppsym}
\VV''_{0,1,2}(\PSZ; \wt A, \wt B)= -\VV''_{0,1,2}(\PSZ; \wt B, \wt A)\, .
\ee
For this reason when we sew a $\wt\HH$ puncture of $\VV''$  to another puncture belonging
to another Riemann surface, it is important to specify which of the two $\wt\HH$ punctures of
$\VV''_{0,1,2}$ is being used for sewing. Following the conventions given below
\refb{estarint},  if the first $\wt\HH$ puncture of
$\VV''_{0,1,2}$ takes part in the sewing, then we shall write $\VV''_{0,1,2}$ to the right of
the $\rstar$ symbol, whereas if the second $\wt\HH$
puncture of $\VV''_{0,1,2}$ takes part in the sewing, 
then we shall write $\VV''_{0,1,2}$ to the left of
the $\lstar$ symbol. Due to this antisymmetry property, the relations of the form 
\refb{eexch2} acquire extra minus sign when one of the subspaces
is $\VV''_{0,1,2}$.

\subsection{Deformation of the interaction terms}

Finally we turn to the interpretation of $\delta\{\Phi^N\}$.
This may be expressed as\footnote{The $(-2\pi i)^{-1}$ factor in \refb{epert1a} arises in
\refb{efifth} from a  $(-2\pi i)^{-1}$ factor included in the definition of $\Omega^{(g,m,n)}_p$
for every additional puncture (see {\it e.g.} \cite{1703.06410}).}
\be \label{efifth}
\delta\{\Phi^N\}=\lambda \, \sum_{g\ge 0} \, \VV'_{g, N+1,0} (\PSZ,  \, \Phi^N )\, ,
\ee
where $\VV'_{g, N+1,0}$ is a $(6g + 2N-4)$  dimensional subspace of $\QQ_{g,N+1,0}$
defined as follows. We begin with the Riemann surfaces associated with the subspace
$\VV_{g,N,0}\subset \QQ_{g,N,0}$ and insert a special puncture at any point on the
Riemann surface outside the unit disks: $|w_i|\ge 1$ for $1\le i\le N$,  where $w_i$ is
the local coordinate around the $i$-th puncture on the Riemann surfaces associated
with $\VV_{g,N,0}$. 
Since we shall always insert the conformally invariant vertex operator $\Psi_0$ at the
special puncture, we do not need to specify the local coordinate at this puncture.
Therefore $\VV'_{g, N+1,0}$ corresponds to a 
subspace of $\QQ_{g,N+1,0}$ modulo the choice of local coordinates at the special
puncture, and all relations involving $\VV'_{g, N+1,0}$ that we shall write below
will be modulo this choice.
Note that for $g=0$ this definition of $\VV'_{g,N+1,0}$ is valid
for $N\ge 3$. We define $\VV'_{0,3,0}$ to be zero. We shall use the convention that
the first puncture of
$\VV'_{g,N+1,0}$ will be the special puncture. From the symmetry of
$\VV_{g,N,0}$ under the permutations of the punctures, it follows that 
$\VV'_{g,N+1,0}$ is
symmetric under the permutations of all the punctures other than the special puncture.
Since $\VV_{g,N,0}$ avoids
regions of the moduli space with separating type degenerations, and since in the
definition of $\VV'_{g,N+1,0}$ we always keep the special puncture away from the other
punctures, $\VV'_{g,N+1,0}$ also avoids separating type degenerations.

Let us define
\be 
\VV'_{N+1,0}\equiv \sum_{g\ge 0} \VV'_{g, N+1,0}\, .
\ee
Then \refb{efifth} may be expressed as
\be \label{exxnn}
\delta\{\Phi^N\}=\lambda \, \VV'_{N+1,0} (\PSZ,  \, \Phi^N )\, .
\ee

The result for $\delta\{A_1\cdots A_N\}$ can be found from \refb{exxnn} by taking $\Phi
=A_1+\cdots+A_N$ and keeping terms proportional to $A_1\cdots A_N$ on both sides. For
example replacing $\Phi$ by $A+\Phi$ and keeping terms linear in $A$ on both sides of
\refb{exxnn} we get
\be \label{exxyy}
\delta\{A\Phi^{N-1}\} = \lambda \, \VV'_{N+1,0} (\PSZ,  A, \Phi^{N-1} )\, .
\ee

In the following analysis we shall make use of various relations involving the
$*$, $\lstar$ and $\rstar$ product of the subspaces $\VV'_{0,2,1}$, $\VV''_{0,1,2}$
and $\VV'_{g,N+1,0}$. Since these vertices are not symmetric in all the punctures, we
need to carefully specify which of the punctures are sewed. In this we shall
use the convention that the special puncture carried by the corresponding Riemann
surfaces never takes part in sewing. Therefore for example when we we have a
subspace of the form $\AAA*\VV'_{g,N+1,0}$, it is the left-most puncture of $\VV'$ 
{\it other
than the special puncture} that takes part in the sewing. Similar convention will be used
for $\VV'_{0,2,1}$ and $\VV''_{0,1,2}$.

The boundary of $\VV'_{N+1,0}$ will play a special role in the subsequent analysis. 
This can be determined as follows. Since 
\be \label{edefdef}
\VV^{\rm deformed}_{N,0}(\Phi^{N})
\equiv \VV_{N,0}(\Phi^N)+\lambda\VV'_{N+1,0}(\PSZ,
\Phi^N)
\ee 
gives the interaction vertex of the deformed theory, it satisfies an identity similar
to \refb{enewrr}:
\be\label{enewrra}
M \, \VV^{\rm deformed}_{M,0}((Q_B+\delta Q_B)\Phi,\Phi^{M-1})
= -{1\over 2}\, \sum_{M_1, M_2\atop
M_1+M_2=M} {M!\over M_1! M_2!}\,  \VV^{\rm deformed}_{M_1+1,0}*' 
\VV^{\rm deformed}_{M_2+1,0}(\Phi^{M_1}|\Phi^{M_2})\, ,
\ee
where $*'$ is defined in the same way as $*$ except that for Ramond sector sewing
we insert $\GG+\delta\GG$ instead of $\GG$ around one of 
the punctures that are sewed.
Now, using \refb{egenpar}, \refb{esewing}-\refb{esewing2} 
and the anti-symmetry property \refb{evppsym}, we can write
\ben \label{eabove}
&& \VV_{M_1+1,0} *' \VV_{M_2+1,0}(\Phi^{M_1}| \Phi^{M_2}) \nonumber \\
&=& \VV_{M_1+1,0} * \VV_{M_2+1,0}(\Phi^{M_1}| \Phi^{M_2})
+ \lambda \, \VV_{M_1+1,0} \rstar \VV''_{0,1,2} \lstar \VV_{M_2+1,0} (\Phi^{M_1}|\PSZ| \Phi^{M_2})\, .
\een
While applying \refb{esewing}-\refb{esewing2}, \refb{egenpar} and \refb{evppsym} 
to arrive at \refb{eabove}, we have
to keep in mind that not all the arguments that appear in the intermediate steps of
the analysis are grassmann even, and therefore there will be
extra signs that have to be computed carefully
using the procedure explained earlier.
Expanding \refb{enewrra} in powers of $\lambda$ using \refb{edefdef},
\refb{eabove},
and collecting the
coefficients of the order $\lambda$ term, we get
\ben
M \, \VV'_{M+1,0}(\PSZ, Q_B\Phi, \Phi^{M-1}) &=&
- \sum_{M_1, M_2\atop
M_1+M_2=M} {M!\over M_1! M_2!}\,  \VV_{M_1+1,0} * 
\VV_{M_2+2,0}'(\Phi^{M_1}|\PSZ, \Phi^{M_2})
\nonumber \\ &&
- {1\over 2} \sum_{M_1, M_2\atop
M_1+M_2=M} {M!\over M_1! M_2!}\, \VV_{M_1+1,0} \rstar \VV''_{0,1,2}
\lstar \VV_{M_2+1,0} (\Phi^{M_1}|\PSZ| \Phi^{M_2})\nonumber \\ &&
- M \, \lambda^{-1}\, \VV_{M,0}(\delta Q_B\Phi, \Phi^{M-1}) \, .
\een
As already mentioned, this relation holds up to the choice of local coordinates at
the special puncture.
Using \refb{esewing2} and \refb{eddqb} we can express this as
\ben \label{exx11}
\p \VV'_{M+1,0}(\PSZ, \Phi^M) &=&
- \sum_{M_1, M_2\atop
M_1+M_2=M} {M!\over M_1! M_2!}\,  \VV_{M_1+1,0} * 
\VV_{M_2+2,0}'(\Phi^{M_1}|\PSZ, \Phi^{M_2})
\nonumber \\ &&
- {1\over 2} \sum_{M_1, M_2\atop
M_1+M_2=M} {M!\over M_1! M_2!}\,
\VV_{M_1+1,0} \rstar \VV''_{0,1,2} \lstar \VV_{M_2+1,0} (\Phi^{M_1}|\PSZ| \Phi^{M_2})
\nonumber \\ &&
- M \, \VV'_{0,2,1}\lstar \VV_{M,0}(\PSZ, \Phi| \Phi^{M-1})\, .
\een
An equivalent relation, that will be useful later, is obtained by
replacing $\Phi$ by $\Phi+A$ and keeping terms linear in $A$ on both sides:
\ben \label{exx12}
M\, \p \VV'_{M+1,0} (\PSZ, A, \Phi^{M-1}) &=& - \sum_{M_1, M_2\atop
M_1+M_2=M} {M!\over M_1! (M_2-1)!}\,  \VV_{M_1+1,0} * 
\VV_{M_2+2,0}'(\Phi^{M_1}|\PSZ, A, \Phi^{M_2-1})
\nonumber \\ && \hskip -1.5in
- \sum_{M_1, M_2\atop
M_1+M_2=M} {M!\over (M_1-1)! M_2!}\,  \VV_{M_1+1,0} * 
\VV_{M_2+2,0}'(A, \Phi^{M_1-1}|\PSZ, \Phi^{M_2}) \nonumber \\ &&
 \hskip -1.5in
- \sum_{M_1, M_2\atop
M_1+M_2=M} {M!\over (M_1-1)! M_2!}\,
\VV_{M_1+1,0} \rstar \VV''_{0,1,2} \lstar \VV_{M_2+1,0} (A, 
\Phi^{M_1-1}|\PSZ| \Phi^{M_2})
\nonumber \\ &&
\hskip -1.5in - M \, \VV'_{0,2,1}\lstar \VV_{M,0}(\PSZ, A| \Phi^{M-1})
- M(M-1) \, \, \VV'_{0,2,1}\lstar \VV_{M,0}(\PSZ, \Phi| A, \Phi^{M-2}).
\nonumber \\
\een

\sectiono{Proof of background independence} \label{s4}

In this section we shall show the existence of $\delta\Phi$ satisfying the $\Delta\simeq 0$ equation,
where $\simeq$ denotes equality up to terms that vanish by leading order equations of motion \refb{eleading}.
We shall seek a solution for $\delta\Phi$ such that for any grassmann even 
state $\wt B\in\wt\HH$:
\be \label{eansatz}
\langle \wt B| c_0^- |\delta\Phi\rangle = \lambda\, \sum_{N\ge 0} 
{1\over N!} \, 
\BB_{N+1,1}(\PSZ, \Phi^N; \wt B)\, ,
\ee
\be \label{edefBB}
\BB_{N+1,1}\equiv \sum_{g\ge 0} \BB_{g, N+1, 1}\, ,
\ee
where $\BB_{g, N+1, 1}$ is a $6g+2N-1$ dimensional subspace of $\QQ_{g,N+1, 1}$
that is to be determined.
The dimension of $\BB_{g,N+1,1}$ is fixed by the requirement that the total ghost number
carried by the arguments minus the dimension of $\BB_{g,N+1,1}$ should be equal to
$6-6g$ due to ghost number conservation.
The first $\wh\HH$ puncture  
of $\BB_{g, N+1, 1}$ is taken to be the special puncture where
we insert the state $\PSZ$. This puncture never takes part in the sewing operation, and
we leave unspecified the choice of local coordinate at this puncture.
$\BB_{g, N+1, 1}$ is taken to be symmetric under the permutations of the rest of the $N$
punctures  in $\wh\HH$ where 
$\Phi$'s are inserted. $\wt B$ is inserted at the only $\wt\HH$ puncture of
$\BB_{g,N+1,1}$. 
We shall look for solutions for $\BB_{g,N+1,1}$ that avoid
separating type degenerations.

If $\wt B$ is grassmann odd then there is an addition minus sign in
\refb{eansatz} arising as follows.
When we convert a grassmann 
odd operator to a grassmann even operator by multiplication by
a grassmann odd c-number $\zeta$, and 
try to strip off $\zeta$ by moving it to the extreme right or left on both sides, we get an 
extra minus
sign arising from the effect of passing $\zeta$ through $c_0^-$ on the left
hand side of \refb{eansatz} when we move $\zeta$ to the right, or from having to
pass $\zeta$ through the $6g+2N-1$ grassmann odd operators 
on the right hand side of \refb{eansatz},
implicit in the definition of $\BB_{g, N+1, 1}(\PSZ, \Phi^N; \wt B)$, when
we move $\zeta$ to the left.

\subsection{Geometrization of the problem of background independence} 

In this section we shall describe the geometric interpretation of different terms appearing
in \refb{eDeltanew}.

\noindent {\bf First term:} 
We begin with the first term on the right hand side of \refb{eDeltanew}. Since $Q_B\wt A$ is
grassmann odd, we have, from \refb{eansatz}, \refb{eipart} and the BRST invariance
of $|\Psi_0\rangle$, that
\ben \label{efirstpre}
&& \langle Q_B\wt A| c_0^-  |\delta \Phi\rangle = - \lambda \, \sum_N {1\over N!} \, 
\BB_{N+1,1}(\PSZ, \Phi^N; Q_B \wt A) \nonumber \\
&=&  \lambda \, \sum_N {1\over N!} \,  \p \BB_{N+1,1}(\PSZ, \Phi^N; \wt A) 
+ \lambda \, \sum_N {1\over (N-1)!} \, \BB_{N+1,1}(\PSZ, \Phi^{N-1}, Q_B\Phi; \wt A)\, .
\een
Using the leading order equations of motion \refb{eleading} this may be rewritten as
\ben \label{efirst}
&& \hskip -.5in \lambda \, \sum_N {1\over N!} \,  \p \BB_{N+1,1}(\PSZ, \Phi^N; \wt A) 
- \lambda \, \sum_{N,M} {1\over (N-1)! M!} \, \BB_{N+1,1}(\PSZ, \Phi^{N-1}, \vp_r; \wt A)
\langle \vp_r^c | c_0^- \GG| \vp_s^c\rangle \langle \vp_s|c_0^- [\Phi^M]\rangle\nonumber\\
&=& \lambda \, \sum_N {1\over N!} \,  \p \BB_{N+1,1}(\PSZ, \Phi^{N}; \wt A) 
- \lambda \, \sum_{N,M} {1\over N! M!} \, \BB_{N+2,1} * \VV_{M+1,0}(\PSZ, \Phi^N; \wt A| 
\Phi^M)\,  ,
\een
where in the second step we have used \refb{egrconf}, \refb{esewing} and \refb{estring} .

\noindent {\bf Second term:} 
In order to give a geometric interpretation of the second term 
 on the right hand side of \refb{eDeltanew}, we express this as
\ben
&& \sum_{N} {1\over N!} \{ (\GG \wt A) \Phi^N \, \vp_r \}  \langle \vp_r^c|c_0^-|
\delta\Phi\rangle \nonumber \\
&=& \lambda\, 
\sum_{N} \sum_M {1\over N!} \, {1\over M!} \, (-1)^{n_s} \, 
\VV_{N+2,0} ((\GG \wt A) , \Phi^N , \vp_r)  \langle \vp_r^c |c_0^- | \vp_s^c\rangle
\, \BB_{M+1,1} (\PSZ, \Phi^M; \vp_s)\, .
\een
Using \refb{esewing1} and the fact that $\dim\, \BB_{g,M+1,1}=6g+2M-1$, we can express this as
\be \label{esecond}
\lambda\, 
\sum_{N} \sum_M {1\over N!} \, {1\over M!} \, \VV_{N+2,0} \rstar \BB_{M+1,1} ((\GG\wt A), \Phi^N |
\PSZ, \Phi^M)\, .
\ee
Note that since $\VV_{N+2,0}$ is fully symmetric in all the punctures we need not specify the
puncture that participates in the sewing. $\BB_{M+2,1}$ is not fully symmetric, but since it has only
one $\wt\HH$ puncture, there is no ambiguity in which puncture takes part in the sewing. 

\noindent {\bf Third term:} 
It follows from \refb{eddqb}
that the third term on the right hand side of \refb{eDeltanew} may be written as
\be \label{ethird}
\langle \wt A|c_0^- 
\delta Q_B |\Phi\rangle
=\lambda \,  \VV'_{0,2,1} (\PSZ,  \Phi; \wt A)\, . 
\ee

\noindent{\bf Fourth term:} Using \refb{egenpar} the fourth term on the right hand side of \refb{eDeltanew} may be expressed as
\ben \label{efourth}
&& \sum_N {1\over N!} \{  \Phi^N \vp_r\} \langle \vp_r^c|c_0^-|\vp_s^c\rangle
 \langle \vp_s| c_0^-  \delta \GG| \wt A\rangle \nonumber \\ 
 &=& \lambda \, \sum_N {1\over N!} (-1)^{n_s+1} 
 \VV_{N+1,0}(\Phi^N, \vp_r) \langle \vp_r^c|c_0^-|\vp_s^c\rangle \VV''_{0,1,2}(\PSZ;
 \wt A, \vp_s)
 \nonumber \\
&=& \lambda\, \sum_N {1\over N!}
 \VV_{N+1,0} \rstar \VV''_{0,1,2} (\Phi^N | \PSZ;\wt A)
\een

\noindent{\bf Fifth term:} We shall now consider the fifth term  on the right hand side of 
\refb{eDeltanew}. Using \refb{exxyy} this may be expressed as
\be \label{ethirdfifth}
\lambda \, \sum_N \, {1\over N!}\, \VV'_{N+2,0} (\PSZ, \, (\GG\, \wt A), \, \Phi^N )
\, .
\ee

\noindent{\bf Sixth term:} Finally the sixth term on the right hand side of \refb{eDeltanew}
may be written as
\be \label{esixth}
-\lambda\, \sum_{N} \, {1\over N!} \, \VV_{N+2,0}(\PSZ, (\GG\wt A), \Phi^N)\, .
\ee

Using \refb{efirst}, \refb{esecond}, \refb{ethird}, \refb{efourth}, \refb{ethirdfifth} and
\refb{esixth} we can now express \refb{eDeltanew} as
\ben
\Delta &=& \lambda\, \sum_N {1\over N!} \,  \p \BB_{N+1,1}(\PSZ, \Phi^N; \wt A) 
- \lambda\, \sum_{N,M} {1\over N! M!} \, \BB_{N+2,1} * \VV_{M+1,0}(\PSZ, \Phi^N; \wt A |  
\Phi^M)
\nonumber \\ &&
+ \lambda\, 
\sum_{N} \sum_M {1\over N!} \, {1\over M!} \, \VV_{N+2,0} \rstar \BB_{M+1,1} ((\GG\wt A), \Phi^N |
\PSZ, \Phi^M) + \lambda\, \VV'_{0,2,1}(\PSZ, \Phi, \wt A)
\nonumber \\ &&
+ \lambda\, \sum_N {1\over N!}
 \VV_{N+1,0} \rstar \VV''_{0,1,2} (\Phi^N | \PSZ; \wt A)
 + \lambda \, \sum_N \, {1\over N!}\, \VV'_{N+2,0} (\PSZ, \, (\GG\, \wt A), \, \Phi^N )
\nonumber \\ &&
-\lambda\, \sum_{N} \, {1\over N!} \, \VV_{N+2,0}(\PSZ, (\GG\wt A), \Phi^N)
\, . 
\een
Demanding the vanishing of $\Delta$ now gives
\ben\label{emaster}
&& \p \BB_{N+1,1}(\PSZ, \Phi^N; \wt A)\nonumber \\
&=& \sum_{M_1,M_2\atop M_1+M_2=N} {N!\over M_1!M_2!} \, \BB_{M_1+2,1} * 
\VV_{M_2+1,0}(\PSZ, \Phi^{M_1};\wt A | \Phi^{M_2})
\nonumber \\ &&
-
\sum_{M_1,M_2\atop M_1+M_2=N} {N!\over M_1!M_2!}  \, \VV_{M_1+2,0} \rstar \BB_{M_2+1,1} ((\GG\wt A), \Phi^{M_1} |
\PSZ, \Phi^{M_2})- \delta_{N,1} \VV'_{0,2,1}(\PSZ, \Phi; \wt A) 
\nonumber \\ &&
-
 \VV_{N+1,0} \rstar \VV''_{0,1,2} (\Phi^N | \PSZ,\wt A)
 -  \VV'_{N+2,0} (\PSZ, \, (\GG\, \wt A), \, \Phi^N )
+ \VV_{N+2,0}(\PSZ, (\GG\wt A), \Phi^N)\, . 
\nonumber \\
\een
Since $\Psi_0$ is inserted at the special puncture, this equation needs to hold as a 
relation between subspaces of $\QQ_{N+1,1}$ up to choice of local coordinates at the
special puncture.

\subsection{Absence of obstruction}

We can solve eq.\refb{emaster} 
for $\BB_{N+1,1}\equiv \sum_{g\ge 0} \BB_{g, N+1,1}$ 
iteratively by carrying out genus expansion on both sides.
Using that fact that $\VV_{0,N,0}$ vanishes for $N\le 2$, it is easy to verify that
the expression for 
$\p\BB_{g, N+1,1}$ obtained from \refb{emaster} contains on the right hand side 
$\BB_{g',N'+1,1}$
for $g'<g$, and / or $\BB_{g, N'+1,1}$ for $N'<N$. 
In particular the equation for $\BB_{0,2,1}$ does not involve any 
$\BB_{g,N,1}$ on the right hand side.
Once this is determined we can solve for $\BB_{0,3,1}$ since its equation involves only
$\BB_{0,2,1}$ on the right hand side. Proceeding this was we can first determine all the
$\BB_{0,N,1}$ iteratively, then determine all the $\BB_{1,N,1}$ and so on.

There is however a possible obstruction to solving these equations. 
Since $\p(\p \BB_{g,N+1,1})=0$, in
order that the equation for $\BB_{g,N+1,1}$ obtained from \refb{emaster} has a solution, we need to show
that the $\p$ annihilates the right hand side of the equation. This gives
\ben\label{eprove}
0 &=&  \sum_{M_1,M_2\atop M_1+M_2=N} {N!\over M_1!M_2!} \, \p\BB_{M_1+2,1} * 
\VV_{M_2+1,0}(\PSZ, \Phi^{M_1};\wt A | \Phi^{M_2})
\nonumber \\ &&
+  \sum_{M_1,M_2\atop M_1+M_2=N} {N!\over M_1!M_2!} \, \BB_{M_1+2,1} * 
\p\VV_{M_2+1,0}(\PSZ, \Phi^{M_1};\wt A | \Phi^{M_2}) \nonumber \\ &&
-
\sum_{M_1,M_2\atop M_1+M_2=N} {N!\over M_1!M_2!}  \, \p\VV_{M_1+2,0} \rstar \BB_{M_2+1,1} ((\GG\wt A), \Phi^{M_1} |
\PSZ, \Phi^{M_2})
\nonumber \\ &&
+
\sum_{M_1,M_2\atop M_1+M_2=N} {N!\over M_1!M_2!}  \, \VV_{M_1+2,0} \rstar \p\BB_{M_2+1,1} ((\GG\wt A), \Phi^{M_1} |
\PSZ, \Phi^{M_2})\nonumber \\ &&
-
 \p\VV_{N+1,0} \rstar \VV''_{0,1,2} (\Phi^N | \PSZ;\wt A) +
 \VV_{N+1,0} \rstar \p\VV''_{0,1,2} (\Phi^N | \PSZ;\wt A) \nonumber \\ &&
  -  \p\VV'_{N+2,0} (\PSZ, \, (\GG\, \wt A), \, \Phi^N )
+ \p\VV_{N+2,0}(\PSZ, (\GG\wt A), \Phi^N)\, .
\een
Since in the iterative scheme described above the right hand side involves $\BB_{g',N',1}$ for 
lower values of $g'$ or $N'$ which already satisfy \refb{emaster},  we need to prove \refb{eprove}
for $\BB_{N'+1,1}$'s appearing on the right hand side 
satisfying \refb{emaster}. This allows us to simplify the different terms on the
right hand side of \refb{eprove} as follows.

The first term on the
right hand side of \refb{eprove} is given by
\ben \label{ei1}
&& I_1\equiv \sum_{M_1,M_2\atop M_1+M_2=N} {N!\over M_1!M_2!} \, \p\BB_{M_1+2,1} * 
\VV_{M_2+1,0}(\PSZ, \Phi^{M_1};\wt A | \Phi^{M_2}) \nonumber \\ &=&
\sum_{M_1,M_2\atop M_1+M_2=N} {N!\over M_1!M_2!} \sum_{M_3,M_4\atop M_3+M_4=M_1}
{M_1!\over M_3! M_4!}
\BB_{M_3+2,1} * 
\VV_{M_4+2,0}* 
\VV_{M_2+1,0}(\PSZ, \Phi^{M_3};\wt A | \Phi^{M_4}|\Phi^{M_2})\nonumber \\ &&
-
\sum_{M_1,M_2\atop M_1+M_2=N} {N!\over M_1!M_2!} \sum_{M_3,M_4\atop M_3+M_4=M_1}
{M_1!\over M_3! M_4!}
\VV_{M_4+1,0}* 
\BB_{M_3+3,1} * 
\VV_{M_2+1,0}(\Phi^{M_4}|\PSZ, \Phi^{M_3};\wt A | \Phi^{M_2})\nonumber \\ && \hskip -.2in
- \sum_{M_1,M_2\atop M_1+M_2=N} {N!\over M_1!M_2!} 
\sum_{M_3,M_4\atop M_3+M_4=M_1}
{M_1!\over M_3! M_4!}
 \, \VV_{M_3+2,0} \rstar \BB_{M_4+2,1} * \VV_{M_2+1,0}((\GG\wt A), \Phi^{M_3} |
\PSZ, \Phi^{M_4}|\Phi^{M_2}) \nonumber \\ && \hskip -.2in
+ \sum_{M_1,M_2\atop M_1+M_2=N} {N!\over M_1!M_2!} 
\sum_{M_3,M_4\atop M_3+M_4=M_1}
{M_1!\over M_3! M_4!}   \BB_{M_4+1,1} \lstar
 \, \VV_{M_3+3,0} * \VV_{M_2+1,0}(\PSZ, \Phi^{M_4}|(\GG\wt A), \Phi^{M_3} |
\Phi^{M_2}) \nonumber \\ &&
- \VV'_{0,2,1} * 
\VV_{N+1,0}(\PSZ; \wt A |\Phi^{N})
\nonumber \\ &&
- \sum_{M_1,M_2\atop M_1+M_2=N} {N!\over M_1!M_2!} \, \VV''_{0,1,2} \lstar
\VV_{M_1+2,0}  * 
\VV_{M_2+1,0}(\PSZ;\wt A |\Phi^{M_1}| \Phi^{M_2}) \nonumber \\ &&
- \sum_{M_1,M_2\atop M_1+M_2=N} {N!\over M_1!M_2!} \, \VV'_{M_1+3,0} * 
\VV_{M_2+1,0}(\PSZ, \GG \wt A, \Phi^{M_1} | \Phi^{M_2})\nonumber \\ &&
+ \sum_{M_1,M_2\atop M_1+M_2=N} {N!\over M_1!M_2!} \, \VV_{M_1+3,0} * 
\VV_{M_2+1,0}(\PSZ, \GG \wt A, \Phi^{M_1} | \Phi^{M_2})\, .
\een
Since this manipulation is somewhat involved, we shall illustrate how the first two terms
on the right hand side of \refb{ei1} arise by picking, in the expression 
for $\p\BB_{M_1+2,1}(\Psi_0, \Phi^{M_1+1};\wt A)$,  
the first term on the right hand side
of \refb{emaster}:
\be \label{eterm}
\sum_{M_3,M_4\atop M_3+M_4=M_1+1} {(M_1+1)!\over M_3!M_4!} \, 
\BB_{M_3+2,1} * 
\VV_{M_4+1,0}(\PSZ, \Phi^{M_3};\wt A | \Phi^{M_4})\, .
\ee
We need to sew this to $\VV_{M_2+1,0}$. This is done by picking a puncture other
than the special puncture in $\p\BB_{M_1+2,1}$ and a puncture of $\VV_{M_2+1,0}$
and sewing them. Now since $\p\BB_{M_1+2,1}$ and $\VV_{M_2+1,0}$ are symmetric
under the permutations of the punctures in $\wh\HH$ 
(other than the special puncture), it does not
matter which puncture we choose for sewing. 
However when we pick the term given in \refb{eterm} in the expression for 
$\p\BB_{M_1+2,1}$,
then we have to allow for the puncture to come either from $\BB_{M_3+2,1}$ or from
$\VV_{M_4+1,0}$ with appropriate weight factors given by $M_3/(M_3+M_4)$ and
$M_4/(M_3+M_4)$ respectively. Thus for example the net contribution to 
$I_1$ from the term where we choose
the sewing puncture of $\p\BB_{M_1+2,0}$ from $\VV_{M_4+1,0}$ will be given by
\be
\sum_{M_1,M_2\atop M_1+M_2=N} {N!\over M_1!M_2!} \sum_{M_3,M_4\atop
M_3+M_4=M_1+1} {(M_1+1)!\over M_3!M_4!} \, {M_4\over M_3+M_4}
\BB_{M_3+2,1} * 
\VV_{M_4+1,0}*\VV_{M_2+1,0}(\PSZ, \Phi^{M_3};\wt A | \Phi^{M_4-1}|\Phi^{M_2})\, .
\ee
By making a change of variable $M_4\to M_4+1$ we recover the first term on the right
hand side of \refb{ei1}. Similarly the result of choosing the sewing puncture in
$\p\BB_{M_1+2,1}$ from $\BB_{M_3+2,1}$ is given by the second term on the right
hand side of \refb{ei1}, with the extra minus sign arising from switching the order of
$\BB_{M_3+2,1}$ and $\VV_{M_4+1,0}$ (see \refb{eexch1}).

The second term on the
right hand side of \refb{eprove} is given by
\ben
&& I_2\equiv  \sum_{M_1,M_2\atop M_1+M_2=N} {N!\over M_1!M_2!} \, \BB_{M_1+2,1} * 
\p\VV_{M_2+1,0}(\PSZ, \Phi^{M_1};\wt A | \Phi^{M_2}) 
\nonumber \\ &=&
- \sum_{M_1,M_2\atop M_1+M_2=N} {N!\over M_1!M_2!} \sum_{M_3,M_4\atop M_3+M_4=M_2} 
{M_2!\over M_3!M_4!}\, \BB_{M_1+2,1} * \VV_{M_3+2,0} * \VV_{M_4+1,0}
(\PSZ, \Phi^{M_1};\wt A | \Phi^{M_3}| \Phi^{M_4}) \, . \nonumber \\
\een
The third term on the
right hand side of \refb{eprove} is given by
\ben 
&& I_3\equiv -
\sum_{M_1,M_2\atop M_1+M_2=N} {N!\over M_1!M_2!}  \, \p\VV_{M_1+2,0} \rstar \BB_{M_2+1,1} ((\GG\wt A), \Phi^{M_1} |
\PSZ, \Phi^{M_2})\nonumber \\ &=&
\sum_{M_1,M_2\atop M_1+M_2=N} {N!\over M_1!M_2!}  \, 
\sum_{M_3,M_4\atop M_3+M_4=M_1} 
{M_1!\over M_3!M_4!} \VV_{M_3+2,0} * \VV_{M_4+2,0} \rstar \BB_{M_2+1,1} 
((\GG\wt A), \Phi^{M_3}| \Phi^{M_4} |
\PSZ, \Phi^{M_2})\nonumber \\ &&
\hskip -.3in 
+ \sum_{M_1,M_2\atop M_1+M_2=N} {N!\over M_1!M_2!}  \, 
\sum_{M_3,M_4\atop M_3+M_4=M_1} 
{M_1!\over M_3!M_4!} \VV_{M_3+1,0} * \VV_{M_4+3,0} \rstar \BB_{M_2+1,1} 
(\Phi^{M_3}| (\GG\wt A), \Phi^{M_4} |
\PSZ, \Phi^{M_2})\, .\nonumber \\
\een
The fourth term on the
right hand side of \refb{eprove} is given by
\ben
&& I_4\equiv \sum_{M_1,M_2\atop M_1+M_2=N} {N!\over M_1!M_2!}  \, \VV_{M_1+2,0} \rstar \p\BB_{M_2+1,1} ((\GG\wt A), \Phi^{M_1} |
\PSZ, \Phi^{M_2}) \nonumber \\ &=& 
\sum_{M_1,M_2\atop M_1+M_2=N} {N!\over M_1!M_2!}  \, 
\sum_{M_3,M_4\atop M_3+M_4=M_2} 
{M_2!\over M_3!M_4!} \, \VV_{M_1+2,0} \rstar \BB_{M_3+2,1} * 
\VV_{M_4+1,0}((\GG\wt A), \Phi^{M_1} |
\PSZ, \Phi^{M_3}|\Phi^{M_4}) \nonumber \\ &&
\hskip -.2in - \sum_{M_1,M_2\atop M_1+M_2=N} {N!\over M_1!M_2!}  \, 
\sum_{M_3,M_4\atop M_3+M_4=M_2} 
{M_2!\over M_3!M_4!} \, \VV_{M_1+2,0} * 
\VV_{M_3+2,0} \rstar \BB_{M_4+1,1} ((\GG\wt A), \Phi^{M_1} |
\Phi^{M_3} |\PSZ, \Phi^{M_4}) \nonumber \\ &&
-N \, \VV_{N+1,0} \rstar \VV'_{0,2,1} ((\GG\wt A), \Phi^{N-1} |
\PSZ, \Phi) \nonumber \\ &&
- \sum_{M_1,M_2\atop M_1+M_2=N} {N!\over M_1!M_2!}  \, \VV_{M_1+2,0} \rstar 
\VV''_{0,1,2} \lstar \VV_{M_2+1,0} ((\GG\wt A), \Phi^{M_1} |
\PSZ | \Phi^{M_2}) \nonumber \\ &&
- \sum_{M_1,M_2\atop M_1+M_2=N} {N!\over M_1!M_2!}  \, 
\VV_{M_1+2,0} * \VV'_{M_2+2,0} ((\GG\wt A), \Phi^{M_1} |
\PSZ, \Phi^{M_2}) \nonumber \\ &&
+ \sum_{M_1,M_2\atop M_1+M_2=N} {N!\over M_1!M_2!}  \, 
\VV_{M_1+2,0} * \VV_{M_2+2,0} ((\GG\wt A), \Phi^{M_1} |
\PSZ, \Phi^{M_2})\, .
\een 
The fifth term on the
right hand side of \refb{eprove} is given by
\ben
&& I_5\equiv - \p\VV_{N+1,0} \rstar \VV''_{0,1,2} (\Phi^N | \PSZ;\wt A) \nonumber \\
&=& \sum_{M_1,M_2\atop M_1+M_2=N} {N!\over M_1!M_2!}  \,
\VV_{M_1+1,0} * \VV_{M_2+2,0} \rstar \VV''_{0,1,2} (\Phi^{M_1}|\Phi^{M_2} 
| \PSZ;\wt A)\, .
\een
Using \refb{evvppboundary}, the sixth term on the
right hand side of \refb{eprove} is given by
\ben
&& I_6\equiv 
 \VV_{N+1,0} \rstar \p\VV''_{0,1,2} (\Phi^N | \PSZ;\wt A)
 \nonumber \\ &=&
\VV_{N+1,0} * \VV'_{0,2,1} (\Phi^N | \PSZ;\wt A)
- \VV'_{0,2,1} \lstar \VV_{N+1,0} (\PSZ, \GG\wt A|\Phi^N)\, .
\een
Using \refb{exx12}, the seventh term on the
right hand side of \refb{eprove} is given by
\ben 
&& I_7\equiv  -  \p\VV'_{N+2,0} (\PSZ, \, (\GG\, \wt A), \, \Phi^N )
\nonumber \\
&=& \sum_{M_1,M_2\atop M_1+M_2=N} {N!\over M_1!M_2!}  \, 
\VV_{M_1+1,0} * \VV'_{M_2+3,0} (\Phi^{M_1}|\PSZ, \GG\wt A, \Phi^{M_2})
\nonumber \\ &&
+  \sum_{M_1,M_2\atop M_1+M_2=N} {N!\over M_1!M_2!}  \, 
\VV_{M_1+2,0} * \VV'_{M_2+2,0}  (\GG\wt A, \Phi^{M_1}|\PSZ,  \Phi^{M_2})
\nonumber \\ &&
+ \sum_{M_1,M_2\atop M_1+M_2=N} {N!\over M_1!M_2!}  \, 
\VV_{M_1+2,0} \rstar \VV''_{0,1,2} \lstar \VV_{M_2+1,0}(\GG\wt A, \Phi^{M_1}|\PSZ|
\Phi^{M_2}) \nonumber \\ &&
+ \VV'_{0,2,1}\lstar \VV_{N+1,0}(\PSZ, \GG\wt A| \Phi^{N})
+ N \, \VV'_{0,2,1}\lstar \VV_{N+1,0}(\PSZ, \Phi| \GG\wt A, \Phi^{N-1})\, .
\een
The eighth term on the
right hand side of \refb{eprove} is given by
\ben
&& I_8 \equiv \p\VV_{N+2,0}(\PSZ, (\GG\wt A), \Phi^N)
\nonumber \\ 
&=& - \sum_{M_1,M_2\atop M_1+M_2=N} {N!\over M_1!M_2!}  \, 
\VV_{M_1+3,0} * \VV_{M_2+1,0} (\PSZ, \GG\wt A, \Phi^{M_1}| \Phi^{M_2})
\nonumber \\ &&
-  \sum_{M_1,M_2\atop M_1+M_2=N} {N!\over M_1!M_2!}  \, 
\VV_{M_1+2,0} * \VV_{M_2+2,0} (\PSZ,  \Phi^{M_1}|\GG\wt A, \Phi^{M_2})\, .
\een

We can simplify some of the expressions by using
\ben
&& \sum_{M_1,M_2\atop M_1+M_2=N} {N!\over M_1!M_2!}  \, 
\sum_{M_3,M_4\atop M_3+M_4=M_1} 
{M_1!\over M_3!M_4!}  f(M_1,M_2,M_3,M_4) \nonumber \\
&=& \sum_{M_2,M_3,M_4\atop M_2+M_3+M_4=N} 
{N!\over M_2! M_3!M_4!}  \, 
f(M_3+M_4,M_2,M_3,M_4)\, ,
\een
etc.
With such rearrangements
we can easily see that the terms on the right hand side of \refb{eprove}
cancel pairwise. In particular:
\begin{enumerate}
\item The first term in $I_1$ cancels  $I_2$.
\item The second term in $I_1$ is anti-symmetric under the exchange $M_2\leftrightarrow
M_4$ due to \refb{eexch1} and vanishes after we sum over $M_2,M_4$.
\item The third term in $I_1$ cancels  the first term in $I_4$.
\item The fourth term in $I_1$ cancels  the second term in $I_3$ after using
\refb{eexch1} and \refb{eexch2}.
\item The fifth term in $I_1$ cancels  the first term in $I_6$ after using
\refb{eexch1}.
\item The sixth term in $I_1$ cancels   $I_5$ after using
\refb{eexch1}, \refb{eexch2} and \refb{evppsym}. 
\item The seventh term in $I_1$ cancels  the first term in $I_7$ after using
\refb{eexch1}.
\item The eighth term in $I_1$ cancels  the first term in $I_8$.
\item The first term in $I_3$ cancels  the second term in $I_4$.
\item The third term in $I_4$ cancels  the fifth term in $I_7$ after using
\refb{eexch2}.
\item The fourth term in $I_4$ cancels  the third term in $I_7$.
\item The fifth term in $I_4$ cancels  the second term in $I_7$.
\item The sixth term in $I_4$ cancels  the second term in $I_8$ after using
\refb{eexch1}.
\item The second term in $I_6$ cancels the fourth term in $I_7$.
\end{enumerate}
This shows that the right hand side of \refb{emaster} has no boundary.

\subsection{Proof of background independence}

Once we have shown that the right hand side of \refb{emaster} has no boundary, one can argue
for the existence
of $\BB_{N+1,1}$ satisfying \refb{emaster} as follows. Let us write \refb{emaster} as
\be \label{einterright}
\p\BB_{N+1,1}(\PSZ, \Phi^N; \wt A) = \bar\VV_{N+1,1}(\PSZ, \Phi^N; \wt A)
-  \bar\VV'_{N+1,1}(\PSZ, \Phi^N;\wt A)\, ,
\ee
where
\ben \label{einterleft}
\bar\VV_{N+1,1}(\PSZ, \Phi^N; \wt A) &=& \VV_{N+2,0}(\PSZ, \GG\wt A, \Phi^N)\, ,
\nonumber \\ 
 \bar\VV'_{N+1,1}(\PSZ, \Phi^N;\wt A) &=& -\sum_{M_1,M_2\atop M_1+M_2=N} {N!\over M_1!M_2!} \, \BB_{M_1+2,1} * 
\VV_{M_2+1,0}(\PSZ, \Phi^{M_1};\wt A | \Phi^{M_2})
\nonumber \\ &&
\hskip -1in +
\sum_{M_1,M_2\atop M_1+M_2=N} {N!\over M_1!M_2!}  \, \VV_{M_1+2,0} \rstar \BB_{M_2+1,1} ((\GG\wt A), \Phi^{M_1} |
\PSZ, \Phi^{M_2})+ \delta_{N,1} \VV'_{0,2,1}(\PSZ, \Phi; \wt A) 
\nonumber \\ &&
+
 \VV_{N+1,0} \rstar \VV''_{0,1,2} (\Phi^N | \PSZ,\wt A)
 +  \VV'_{N+2,0} (\PSZ, \, (\GG\, \wt A), \, \Phi^N )\, .
 \een
 Both $ \bar\VV_{N+1,1}$ and $\bar\VV'_{N+1,1}$ are appropriate subspaces of 
 $\QQ_{N+1,1}$ modulo the choice of local coordinates at the special puncture where
 $\Psi_0$ is inserted. Vanishing of the right hand side of \refb{eprove}
implies that $\bar \VV_{N+1,1}$ and $\bar\VV'_{N+1,1}$ have common boundary. 
Furthermore since by assumption the $\BB_{g,M,1}$'s appearing on the right hand side
of \refb{einterleft} do not contain any separating type degenerations, $\bar\VV'_{N+1,0}$
also does not contain any separating type degeneration.
Now noting that the dimensions of $\bar \VV_{g,N+1,1}$ and $\bar\VV'_{g,N+1,1}$ are both
given by $6g-2+2N$, their projections on to the base of $\QQ_{g,N+1,1}$
will generically be $6g-2+2N$ dimensional
-- the same as the dimension of the base given by the moduli space of Riemann surfaces of
genus $g$ with $N+2$ punctures.
Therefore the interiors of $\bar \VV_{g,N+1,1}$ and $\bar\VV'_{g,N+1,1}$
either both have the same projection into the moduli space, or they are complements of
each other and have opposite orientation. It is easy to argue that the former is true -- since
neither $\bar \VV_{g,N+1,1}$ nor $\bar\VV'_{g,N+1,1}$ contain any Riemann surface with
separating type degenerations, they cannot be complements of each other. This shows that 
we can
regard $\bar \VV_{N+1,1}$ and $\bar\VV'_{N+1,1}$ as two different section segments with the
same boundary.
Therefore
if we can show that $\bar \VV_{N+1,1}$ and $\bar\VV'_{N+1,1}$ are in the same
`homology class' of $\QQ_{g,N+1}$, it will establish the
existence of a $\BB_{N+1,1}$ that has $\bar \VV_{N+1,1}-\bar\VV'_{N+1,1}$ 
as its boundary.
If we forget the fiber coordinates associated with PCO locations, then it  
follows from an analysis
described in 
\cite{back1} that
$\bar \VV_{N+1,1}$ and $\bar\VV'_{N+1,1}$ are deformable to each other and are therefore
in the same homology class. 
Therefore we need to focus on the part of
the fiber  parametrized by the choice of PCO locations. Strictly speaking this does not describe 
a regular manifold since we need to exclude codimension two subspaces to 
avoid spurious poles.
Nevertheless the rules for integration along these fiber directions -- called 
vertical integration\cite{1408.0571,1504.00609} 
-- are such
that they behave {\it as if} these fiber directions describe a homologically 
trivial space -- any
closed submanifold may be regarded as the boundary of another manifold. 
This may seem
counterintuitive -- for example given two possible locations of a PCO on a Riemann surface, one could
have multiple curves that interpolate between these two points which differ by non-contractible cycles
of the Riemann surface and are therefore not deformable to each other. However the integrals of
$\Omega_p$ along
all of these curves give the same result since the result of vertical integration only depends on the
end-points and not on the details of the path. Therefore these paths are equivalent. 
This may give the impression that the integral of $\Omega_p$ along any closed 
vertical subspace vanishes, but this is not quite true. For example if we have a pair
of PCO's and consider the one dimensional 
vertical cycle in which their locations are moved as
\be 
(z_1,z_2)\to (z_1', z_2) \to (z_1', z_2') \to (z_1, z_2') \to (z_1, z_2)\, ,
\ee
then integral of $\Omega_1$ along this cycle does not vanish. Nevertheless we can
effectively reexpress this as an integral of $d\Omega_1$ along a two dimensional
vertical cycle (or more precisely a similar integral where $d\Omega_1$ is replaced
by $\Omega_{2}$ with modified argument using \refb{euse1}).
The net result is that 
one can regard the fiber directions of $\wt\PP_{g,m,n}$ labelled by the  
PCO locations to be homologically trivial\cite{1504.00609}. 
This in turn establishes the existence of
$\BB_{N+1,1}$ satisfying \refb{einterright}.

This analysis also gives an iterative proof that the subspaces $\BB_{g,N+1,1}$ do not include separating 
type degenerations. Assuming that the $\BB_{g,M+1,1}$'s appearing on the right hand side of 
\refb{einterleft}
do not involve separating type degenerations, and using the knowledge that none of the subspaces
$\VV_{M,0}$, $\VV'_{M,0}$, $\VV'_{0,2,1}$ and $\VV''_{0,1,2}$ contain separating type degenerations,
we see that $\bar\VV_{N+1,1}$ and $\bar\VV'_{N+1,1}$ defined in \refb{einterleft}
do not contain separating type degenerations.
Therefore $\BB_{N+1,1}$ computed from \refb{einterright} 
can also be chosen to avoid separating type
degenerations.

Finally, we note that the choice of $\BB_{N+1,1}$ satisfying \refb{einterright} is not unique. 
Since $\QQ_{g,N+1,1}$ is
an infinite dimensional space while $\BB_{g,N+1,1}$ is finite dimensional, 
there are in general
infinite families of subspaces of $\QQ_{N+1,1}$ with the same boundary given by the
right hand side of \refb{einterright}.
Therefore the field redefinition relating the two string field theories
is not unique. This is simply a reflection of the infinite parameter gauge
invariance of the theory as mentioned in the introduction. 

\sectiono{Generalizations to other theories} \label{sgen}

The extension of the analysis given above to type II superstring field theory is
straightforward. The 1PI equations of motion of the interacting fields still take the
form given in \refb{einteom}, with the only difference that now the string field
contains four different sectors: NSNS, NSR, RNS and RR. The operator $\GG$
in these four sectors take the form 1, $\XX_0$, $\bar\XX_0$ and $\XX_0\bar\XX_0$
respectively, where now $\bar\XX_0$ represents the zero mode of the PCO in the
left-moving sector of the world-sheet theory. 
The analog of \refb{eope1} now takes the form
\ben \label{eope11}
&& T_F(z) \, \OO(w,\bar w) \simeq - (z-w)^{-1} \, \overline{\OO}(w,\bar w) 
+ \hbox{less singular terms}
\nonumber \\ &&
\bar T_F(\bar z) \, \OO(w,\bar w) \simeq (\bar z-\bar w)^{-1} \, \wh\OO(w,\bar w) 
+ \hbox{less singular terms}
\nonumber \\ &&
T_F(z) \, \wh\OO(w,\bar w) \simeq (z-w)^{-1} \, \wt \OO(w,\bar w) 
+ \hbox{less singular terms}\nonumber \\ &&
\bar T_F(\bar z) \, \overline{\OO}(w,\bar w) \simeq (\bar z-\bar w)^{-1} \, \wt\OO(w,\bar w) 
+ \hbox{less singular terms}\nonumber \\ &&
T_F(z) \, \overline{\OO}(w,\bar w) \simeq - {1\over 4} (z-w)^{-2} \OO(w,\bar w) 
- {1\over 4} (z-w)^{-1} \p\OO(w, \bar w)\
+ \hbox{less singular terms}\nonumber \\ &&
\bar T_F(z) \, \wh\OO(w,\bar w) \simeq {1\over 4} (\bar z-\bar w)^{-2} 
{\OO}(w,\bar w) + {1\over 4} (\bar z-\bar w)^{-1} \bar\p {\OO}(w, \bar w)\
+ \hbox{less singular terms} \nonumber \\ &&
T_F(z) \, \wt\OO(w,\bar w) \simeq {1\over 4} (z-w)^{-2} \wh\OO(w,\bar w) + {1\over 4} (z-w)^{-1} \p\wh\OO(w, \bar w)\
+ \hbox{less singular terms}\nonumber \\ &&
\bar T_F(z) \, \wt\OO(w,\bar w) \simeq {1\over 4} (\bar z-\bar w)^{-2} 
\overline{\OO}(w,\bar w) + {1\over 4} (\bar z-\bar w)^{-1} \bar\p\overline{\OO}(w, \bar w)\
+ \hbox{less singular terms} \nonumber \\
\, ,
\een
where $\OO$ is a dimension $(1/2,1/2)$ operator. The perturbation \refb{epert1a}
retains the same form
\be \label{epert1b}
 - {\lambda\over 2\pi i} \,\int dz\wedge d\bar z \, \wt \OO(z,\bar z)\, .
\ee
Eqs.\refb{e3.6} and \refb{e3.7} are replaced by
\be \label{e3.6r}
\delta Q_B = \lambda\, \ointop d\bar z \, c(z) \, \wt\OO(z,\bar z) +  \lambda\, 
\ointop d z \, \bar c(\bar z) \, \wt\OO(z,\bar z) +{1\over 4} \lambda\,  \ointop d\bar z \,
\, \gamma(z) \, \wh \OO(z,\bar z)
+ {1\over 4} \lambda\,  \ointop d z \,
\, \bar\gamma(\bar z) \, \overline{\OO}(z,\bar z)
\, ,
\ee
and
\ben\label{e3.7r}
\delta \XX_0 &=&{1\over 4}\, 
\lambda\,  \ointop d\bar z \, z^{-1} \, e^{\phi}(z) \, \wh\OO(z,\bar z)\, , \nonumber \\
\delta \bar\XX_0 &=&{1\over 4}\, 
\lambda\,  \ointop d z \, \bar z^{-1} \, e^{\bar\phi}(\bar z) \, \overline{\OO}(z,\bar z)\
\, .
\een
Besides the identities \refb{ennxx}, we now also have
\be \label{ennxx2}
[Q_B, \delta\bar \XX_0]+[\delta Q_B, \bar \XX_0]=\OO(\lambda^2), \quad
[\XX_0, \delta \bar\XX_0]=\OO(\lambda^2), \quad 
[\bar\XX_0, \delta \XX_0]=\OO(\lambda^2)\, .
\ee

The proof of background independence of the equations of motion proceeds as 
in the case of heterotic string theory with the only difference that we need to pay
a little more attention to the definition of $\VV''_{0,1,2}$. It is defined to be zero
in the NSNS sector as for the NS sector of the
heterotic string theory, whereas its definition in the NSR and RNS sectors is
similar to that for
the R sector of the heterotic string theory. The definition of 
$\VV''_{0,1,2}\subset \QQ_{0,1,2}$ in the RR sector involves a vertical
segment in which the PCO arrangement moves from $\XX_0\bar\XX_0$ around
$\infty$ to $\XX_0\bar\XX_0$ around 0. We can do this either by first moving $\bar\XX_0$
from $\infty$ to 0 and then $\XX_0$ from $\infty$ to 0 or vice versa. Since 
$\XX_0$ and $\delta \bar\XX_0$ commute and $\bar\XX_0$ and $\delta\XX_0$
commute, the two ways of moving the PCO's from $\infty$ to 0 give identical
results for $\VV''_{0,1,2}$. The rest of the analysis is identical to
that for  the heterotic string theory.

One could also explore the possibility of extending the analysis to other versions
of superstring field theory.
There are various other versions of open and closed superstring field 
theories at tree level. These theories
may be broadly divided into two classes -- those based on $A_\infty$ algebra (for open string) or 
$L_\infty$ algebra (for
closed string), and those that do not have manifest underlying $A_\infty$ or $L_\infty$ algebra. The
former class of theories  have the property that 
the sum of the Feynman diagrams of the theory produces in a 
straightforward  manner the string amplitudes computed from the first 
quantized formalism.
The tree level 
open and closed superstring field theories constructed in 
\cite{1312.2948,1403.0940,1506.05774,1602.02582,1602.02583,1703.08214}. provide 
sample examples of theories in this class.
In the second class of theories, which include for example the theories described in
\cite{0109100,0406212,0409018,1508.00366},
the Feynman diagrams do 
not produce the expected string amplitudes in a
straightforward manner, and so far this has been checked either by explicit computation 
in a case by case
basis (see {\it e.g.} \cite{9912120,1312.1677,1612.00777}), 
or by showing its equivalence to another version of the theory that uses $A_\infty/L_\infty$
algebra after suitable gauge fixing and field redefinition\cite{1505.01659,1505.02069,1510.00364}.
We expect that the method described in this paper can be used to prove the background independence
of the first class of theories  in a straightforward manner. However proving 
background independence of other versions of string field theory may require developing new
techniques. 

\bigskip

\noindent{\bf Acknowledgments:}  
I wish to thank Barton Zwiebach for useful discussions and comments on an
earlier version of the manuscript. 
A preliminary version of this work was reported at the SFT@HIT
conference at the Holon Institute of Technology, Tel Aviv.
I wish to thank the participants at the workshop for lively discussions.
I acknowledge 
the hospitality of the
Kavli Institute for Theoretical Physics  where part of the
work was done. 
This work was supported
in part by the JC Bose fellowship of the Department of Science and Technology,
India.

\end{document}